\documentclass[10pt,journal,compsoc]{IEEEtran}
\ifCLASSOPTIONcompsoc
  \usepackage[nocompress]{cite}
\else
  \usepackage{cite}
\fi
\ifCLASSINFOpdf
\else
\fi
\usepackage{amsmath,amssymb,amsfonts}
\usepackage{algpseudocode}
\usepackage[linesnumbered,ruled,vlined]{algorithm2e}
\usepackage{pifont}
\usepackage{graphicx}
\usepackage{textcomp}
\usepackage{xcolor}
\usepackage{pgf}
\usepackage{tikz}
\usepackage{color}
\usepackage{stfloats}
\usepackage{multirow}
\usepackage{diagbox}
\usepackage[hyphens]{url}

\usepackage{fancyhdr}
\usepackage[normalem]{ulem}

\usepackage{balance}
\usepackage{mathptmx} 

\usepackage{float}
\usepackage{booktabs}

\usepackage{comment}
\usepackage[bookmarks=true,breaklinks=true,colorlinks,linkcolor=black,citecolor=blue,urlcolor=black]{hyperref}

\newcommand{\squishlist}{
   \begin{list}{$\bullet$}
    { \setlength{\itemsep}{0pt}      \setlength{\parsep}{0pt}
      \setlength{\topsep}{3pt}       \setlength{\partopsep}{0pt}
      \setlength{\listparindent}{-2pt}
      \setlength{\itemindent}{-5pt}
      \setlength{\leftmargin}{1em} \setlength{\labelwidth}{0em}
      \setlength{\labelsep}{0.5em} } }

\newcommand{\squishend}{
    \end{list}  }

\newcommand*\blackcircledempty[1]{\tikz[baseline=(char.base)]{
        \node[shape=circle, text={rgb,255:red,0;green,0;blue,0}, font=\small, draw={rgb,255:red,0;green,0;blue,0},inner sep=0.5pt] (char) {#1};}}

\begin{document}
%
\title{HiHGNN: Accelerating HGNNs through Parallelism and Data Reusability Exploitation}
%
%
%
%

\author{Runzhen~Xue,
        Dengke~Han,
        Mingyu~Yan,~\IEEEmembership{Member,~IEEE}, \\
        Mo~Zou, 
        Xiaocheng~Yang,
        Duo~Wang,
        Wenming~Li, 
        Zhimin~Tang, \\
        John~Kim, \IEEEmembership{Senior~Member,~IEEE},
        Xiaochun~Ye,
        and~Dongrui~Fan,
        ~\IEEEmembership{Senior~Member,~IEEE}
\IEEEcompsocitemizethanks{
\IEEEcompsocthanksitem 
Runzhen Xue, Dengke Han, Mingyu Yan, Duo Wang, Wenming Li, Zhimin Tang, Xiaochun Ye, and Dongrui Fan are with the State Key Lab of Processors, Institute of Computing Technology, Chinese Academy of Sciences, Beijing 100045,
China, and also with the University of Chinese Academy of Sciences, Beijing
101408, China. 
Mingyu Yan is also with the Shanghai Innovation Center for Processor Technologies, Shanghai 201203, China.
E-mail: \{xuerunzhen21s, handengke21s, yanmingyu, wangduo18z, liwenming, tang, yexiaochun, fandr\}@ict.ac.cn. 

\IEEEcompsocthanksitem 
Mo Zou and Xiaocheng Yang are with the State Key Lab of Processors, Institute of Computing Technology, Chinese Academy of Sciences, Beijing 100045,
China
E-mail: \{zoumo, yangxiaocheng\}@ict.ac.cn. 
 
\IEEEcompsocthanksitem 
John Kim is with the School of Electrical Engineering, Korea Advanced Institute of Science and Technology, Daejeon 305701, Korea. E-mail: jjk12@kaist.ac.kr.

\IEEEcompsocthanksitem 
Runzhen Xue and Dengke Han contributed equally to this work. Mingyu Yan is the corresponding author of this paper.

}


}

%
%

\markboth{}
{Shell \MakeLowercase{\textit{et al.}}: Bare Demo of IEEEtran.cls for Computer Society Journals}
%



\IEEEtitleabstractindextext{%
\begin{abstract}

Heterogeneous graph neural networks (HGNNs) have emerged as powerful algorithms for processing heterogeneous graphs (HetGs), widely used in many critical fields.
To capture both structural and semantic information in HetGs, HGNNs first aggregate the neighboring feature vectors for each vertex in each semantic graph and then fuse the aggregated results across all semantic graphs for each vertex. 
Unfortunately, existing graph neural network accelerators are ill-suited to accelerate HGNNs. This is because they fail to efficiently tackle the specific execution patterns and exploit the high-degree parallelism as well as data reusability inside and across the processing of semantic graphs in HGNNs.

In this work, we first quantitatively characterize a set of representative HGNN models on GPU to disclose the execution bound of each stage, inter-semantic-graph parallelism, and inter-semantic-graph data reusability in HGNNs.
Guided by our findings, we propose a high-performance HGNN accelerator, HiHGNN, to alleviate the execution bound and exploit the newfound parallelism and data reusability in HGNNs.
Specifically, we first propose a bound-aware stage-fusion methodology that tailors to HGNN acceleration, to fuse and pipeline the execution stages being aware of their execution bounds. Second, we design an independency-aware parallel execution design to exploit the inter-semantic-graph parallelism. Finally, we present a similarity-aware execution scheduling to exploit the inter-semantic-graph data reusability.
Compared to the state-of-the-art software framework running on NVIDIA GPU T4 and GPU A100, HiHGNN respectively achieves an average 40.0$\times$ and 8.3$\times$ speedup as well as 99.59\% and 99.74\% energy reduction with quintile the memory bandwidth of GPU A100.


\end{abstract}

\begin{IEEEkeywords}
Heterogeneous graph neural network, Graph neural network, HGNN accelerator, GNN accelerator, HGNN, GNN.
\end{IEEEkeywords}
}

\maketitle

\IEEEdisplaynontitleabstractindextext

%
\IEEEpeerreviewmaketitle






\section{Introduction}\label{sec:intro}

\IEEEPARstart{M}{any} real-world data in complex systems are naturally represented as heterogeneous graphs (HetGs), which possess not only structural information but also rich semantic information~\cite{HG_survey}.
HetGs consist of multiple types of entities and relations which are embodied by various types of vertices and edges, respectively.
This is the major difference of that with homogeneous graphs (HomoGs), which contain only a single type of vertices and edges and thus only represent structural information.
Due to the powerful representation ability of HetG, it has been widely adopted in many critical fields such as knowledge graph~\cite{oh2018knowledge, socher2013reasoning, zhang2019iteratively, chen2017task}, social network~\cite{yasunaga2019scisummnet,zhou2015cross,tajeuna2018modeling,zheng2020clustering}, and many others.


Heterogeneous graph neural networks (HGNNs) originate from the insufficiency of using graph neural networks (GNNs) to process HetGs' semantic information.
Designed for HomoGs to capture their structural rather than semantic information, GNNs recursively aggregate the feature vectors of neighboring vertices~\cite{GCN,comprehensive_gnn_survey,Comprehensive_Survey_GNN_Distributed_Training} to generate the final embedding vector for each vertex.
In contrast, HGNNs use a different execution semantic to capture both pieces of information.
They usually can be partitioned into four major execution stages~\cite{R-GCN,R-GAT,HAN,Simple-HGN,SeHGNN}:
\blackcircledempty{1} \textit{Semantic Graph Build} (SGB) stage partitions the original HetG into several semantic graphs; 
\blackcircledempty{2} \textit{Feature Projection} (FP) stage transforms the feature vector of each vertex in each semantic graph to a new one using a multi-layer perceptron (MLP); 
\blackcircledempty{3} \textit{Neighbor Aggregation} (NA) stage aggregates features from neighbors for each vertex in each semantic graph; 
\blackcircledempty{4} \textit{Semantic Fusion} (SF) stage fuses semantic information revealed by all semantic graphs, i.e., fuses the results of the NA stage across different semantic graphs for each vertex.
HGNNs have achieved excellent prediction accuracy in the processing of HetG and become at the heart of a broad range of critical fields~\cite{hgnn_survey_tangjie,hgnn_survey_hanjiawei,hgnn_survey_shichuan,hgnn_survey_shiruipan} such as recommendation systems~\cite{li2022disentangled,fan2019metapath}, medical analysis~\cite{luo2021imas}, knowledge inference~\cite{CompGCN,A2N,M2GNN}, malicious account detection~\cite{liu2018heterogeneous}, information retrieval~\cite{mao2020item}, shop search~\cite{niu2020dual}, etc.


Deriving from the above workflow, HGNNs exhibit different performance bottlenecks and new acceleration opportunities, compared with GNNs.
According to our quantitative characterization, different stages of HGNNs face different execution bounds, causing low utilization of compute and memory components. 
In addition, HGNNs earn the opportunity to exploit the high-degree parallelism and data reusability in the parallel processing of semantic graphs, which brings the possibility to improve overall performance and efficiency.

Unfortunately, existing GNN accelerators lack the HGNN-oriented design and optimization to efficiently accelerate HGNNs. In particular, a stage-fusion methodology specific to the execution semantics of HGNNs is missed in their designs~\cite{HyGCN,awbgcn}, to improve the utilization of hardware components. Besides, they fail to efficiently exploit the high-degree inter-semantic-graph parallelism and data reusability in the processing of HGNNs.



In this work, we quantitatively characterize a set of representative HGNN models on GPU to ferret out the execution patterns, performance bottlenecks, and acceleration opportunities of HGNNs.
Guided by the characterizations, we design a high-performance HGNN accelerator, called HiHGNN, to exploit the newfound parallelism and data reusability in HGNNs.
Specifically, 
we first propose a novel stage-fusion methodology tailoring to HGNNs, called bound-aware stage fusion. It decomposes and reorganizes coarse-grained stages of HGNNs into fine-grained stages with explicit execution bounds, enabled by a stage-fusion programming model. Then, these fine-grained stages are fused and allowed pipelined execution through a stage-fusion hardware datapath, ensuring higher compute and memory utilization.
Second, we propose an independency-aware parallel execution design to exploit the inter-semantic-graph parallelism. 
This design is built by a multi-lane architecture equipped with workload-aware scheduling, which exploits the high-degree parallelism exposed by the independencies between the processing of semantic graphs, thereby additional hardware resources can be added to further improve performance.
Third, we propose a similarity-aware execution scheduling to harness the potential inter-semantic-graph data reusability. The degree of data reusability between semantic graphs is proportionate to their similarity. To leverage this insight, we construct a hypergraph by taking each semantic graph as a vertex and the similarity between them as edge weights. By conducting the shortest Hamilton path algorithm on this hypergraph, we generate an execution order for the semantic graphs to maximally exploit the inter-semantic graph data reusability.

To summarize, we list our contributions as follows: \par
\squishlist
  \item
  We conduct a quantitative characterization of HGNNs on GPU to uncover the execution patterns, performance bottlenecks, and acceleration opportunities of HGNNs.

  \item
  We propose a high-performance HGNN accelerator, called HiHGNN, to alleviate performance bottlenecks and exploit the newfound parallelism and data reusability in HGNNs. 
  
  \item 
  We propose a bound-aware stage fusion methodology that tailors to HGNN acceleration, including a novel programming model and hardware datapath. Moreover, we propose an independency-aware parallel execution and a similarity-aware execution scheduling to respectively exploit the inter-semantic-graph parallelism and data reusability.
  
  \item
  We implement HiHGNN in RTL and evaluate it through a detailed microarchitectural simulation and on FPGA. 
  Compared to the state-of-the-art software framework running on NVIDIA GPU T4 and GPU A100, HiHGNN achieves an average 40.0$\times$ and 8.3$\times$ speedup as well as 99.59\% and 99.74\% energy reduction, respectively.

\squishend

\section{Background}\label{sec:background}
\begin{figure*}[!htbp] 
	\vspace{-5pt}
	\centering
	\includegraphics[width=0.98\textwidth]{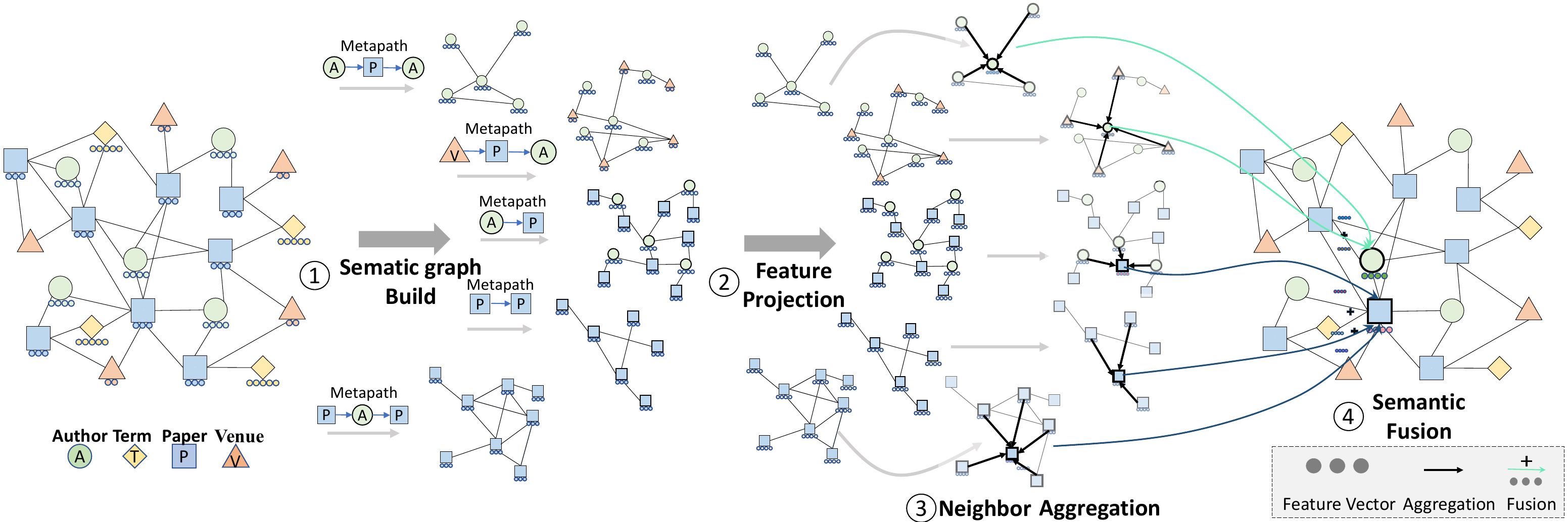}
	\vspace{-10pt}
	\caption{Illustration of HGNNs.}
	\label{fig:HGNN}
	\vspace{-5pt}
\end{figure*}

In this section, we introduce the relevant concepts for HGNNs using Table~\ref{tb:notation}, Fig.~\ref{fig:HGNN}, Algorithm \ref{alg:Original_Programming_Model}, and Table~\ref{tab:HGNN_general_framework}.

\textbf{Heterogeneous Graph.} 
A HetG shown in Fig.~\ref{fig:HGNN}, is defined as $G=(V,E,\mathcal{T}^v,\mathcal{T}^e)$~\cite{SeHGNN,Simple-HGN} using notations in Table~\ref{tb:notation}, where $V$ is the set of vertices with a vertex type mapping function $\phi:V\rightarrow\mathcal{T}^v$, and $E$ is the set of edges with an edge type mapping function $\psi:E\rightarrow\mathcal{T}^e$.
Each vertex $v_i{\in}V$ is attached with a vertex type $c_v{=}\phi(v_i){\in}\mathcal{T}^v$. Each edge $e_{u,v}{\in}E\,$ is attached with a relation $r_{c_v,c_u}{=}\psi(e_{u,v}){\in}\mathcal{T}^e$, starting from the source vertex $u$ to the target vertex $v$.  A graph is heterogeneous when $|\mathcal{T}^v|+|\mathcal{T}^e|>2$, otherwise it is homogeneous.

\textbf{Semantic Graph.} The semantic graph is generated by different metapath. 
A metapath refers to a sequence of vertex types and edge types that captures specific semantic relationships between vertices. It provides a higher-level abstraction of the graph structure and enables the identification of meaningful paths and patterns within the graph. By defining and utilizing metapaths, we can gain insights into the complex relationships and dependencies present in HetGs.
For example in Fig.\ref{fig:HGNN}, a HetG is partitioned into several semantic graphs based on metapaths~\cite{sun2011pathsim,sun2012mining}, such as author$\rightarrow$paper$\rightarrow$author (abbreviated as APA) that represents a coauthor relationship.

\begin{table}[!t]
\centering
\caption{Notations and corresponding explanations.}
\label{tb:notation}
\resizebox{0.48\textwidth}{!}{
\tabcolsep=0.5pt
\begin{tabular}{cc|cc}
\toprule
Notation                & Explanation                           & Notation                  & Explanation       \\ \midrule
$G$                     & heterogeneous graph                   & $V$                       & vertex set \\
$E$                     & edge set                              & $\mathcal{T}^v$           & vertex type set \\
$\mathcal{T}^e$         & edge type set                         & $u,\,v$                   & vertex \\
e ($e_{u,v}$)          & edge (from $u$ to $v$)                & $G^{\mathcal{P}}$ & semantic graph \\
$r$, $\mathcal{P}$          & relation, metapath   & c ($c_v$)                 & vertex type \\
$h$                     & vertex or relation embedding                      & $\mathcal{N}_v$           & neighbor set of vertex $v$ \\
$W$                     & transformation weight matrix          & $h'$                      & projected vertex feature \\
$a$                     & attention vector                      & $||$                      & concatenation \\
$\alpha_{u,v}$          & attention importance    & $\theta (\theta^{\mathcal{P}}_{u,v})$ & attention coefficient \\
$z$                     & intermediate aggregation feature      & $b$                       & transformation bias \\
$\sigma$                & non-linear functions                  & $x$                       & original vertex feature \\

\bottomrule
\vspace{-15pt}
\end{tabular}}
\end{table}

\begin{algorithm}[!t]
    \SetAlgoLined
    \label{alg:Original_Programming_Model}
    \footnotesize
    \caption{\textbf{HGNN Programming Model}}
    \For{each vertex $v$}{
        $h_{v}'$=Feature Projection($h_{v}$)\;
    }
    \For{each semantic graph $G^{\mathcal{P}}$}{
        \For{each vertex pair ($u,v$)}{
            $z_{v}^\mathcal{P}$=Neighbor Aggregation($h_{u}'$)\;
        }
    }
    \For{each vertex $v$}{
        \For{each semantic graph $G^{\mathcal{P}}$}{
            $z_v$=Semantic Fusion($z_{v}^\mathcal{P}$)\;
        }
    }
    
\end{algorithm}

\begin{table*}[!ht]
\small\centering
\caption{The illustration of representative HGNN models.}
\label{tab:HGNN_general_framework}
\resizebox{1.0\textwidth}{!}{
    \begin{tabular}{|c|ccc|c|}
    \hline
     &
      \multicolumn{1}{c|}{HAN~\cite{HAN}} &
      \multicolumn{1}{c|}{R-GAT~\cite{R-GAT}} &
      R-GCN~\cite{R-GCN} &
      S-HGN~\cite{Simple-HGN} \\ \hline
      
    \begin{tabular}[c]{@{}c@{}}Feature\\ Projection\end{tabular} &
      \multicolumn{1}{c|}{${h}'_v=W^{c_v} {x}_v$} &
      \multicolumn{2}{c|}{${h}_v^r=W^{r} {x}_v$} &
      ${h}'_v=W^{c_v} {x}_v$ \\ \hline
    \begin{tabular}[c]{@{}c@{}}Neighbor\\ Aggregation\end{tabular} &
      \multicolumn{2}{c|}{\begin{tabular}[c]{@{}c@{}}$\theta^\mathcal{P}_{u,v}=\sigma({a}^T_{\mathcal{P}}\cdot [{h}'_u || {h}'_v])$,\\ $\alpha^\mathcal{P}_{u,v}=\frac{\exp(\theta^\mathcal{P}_{u,v})}{\sum_{k\in\mathcal{N}^\mathcal{P}_v}\exp(\theta^\mathcal{P}_{v,k})}, {z}^\mathcal{P}_v=\sigma(\sum_{u\in\mathcal{N}^\mathcal{P}_v}\alpha^\mathcal{P}_{u,v}{h}'_u)$\end{tabular}} &
      ${z}_v^r=\frac{1}{|\mathcal{N}_v^r|}\sum_{u\in\mathcal{N}_v^r}{h}^r_u$ &
      \begin{tabular}[c]{@{}c@{}}$\alpha_{u,v}=\frac{\exp \left(\sigma\left(\boldsymbol{a}^{T}\left[{h}'_{u}\left\|\boldsymbol{h}'_v\right\| \boldsymbol{W}_{r} \boldsymbol{h}_r\right]\right)\right)}{\sum_{k \in \mathcal{N}_{v}} \exp \left(\sigma\left(\boldsymbol{a}^{T}\left[{h}'_k\left\|{h}'_v\right\| W_{r} \boldsymbol{h}_r\right]\right)\right)}$,\\ ${h}_{v} =\sum_{u \in \mathcal{N}_{v}^r} \alpha_{u,v} {h}'_{u}$\end{tabular} \\ \hline
    \begin{tabular}[c]{@{}c@{}}Semantic\\ Fusion\end{tabular} &
      \multicolumn{1}{c|}{\begin{tabular}[c]{@{}c@{}}$w_{\mathcal{P}}=\frac{1}{|V^{\mathcal{P}}|}\sum_{v\in V^{\mathcal{P}}}{q}^T\cdot\tanh({W}^{\mathcal{P}}{z}_v^\mathcal{P}+{b})$,\\ $\beta_{\mathcal{P}_i}=\frac{\exp(w_{P_i})}{\sum_{\mathcal{P}_j}\exp(w_{P_j})},\,\,{h}_v = \sum_{\mathcal{P}_i}\beta_{\mathcal{P}_i}{z}^{\mathcal{P}_i}_v$\end{tabular}} &
      \multicolumn{1}{c|}{${h}_v =\frac{1}{|\mathcal{P}|} \sum_{\mathcal{P}_i}{z}^{\mathcal{P}_i}_v$} &
      ${h}_v=\sum_r {z}_v^r+W^{c_v}{x}_v$ &
      --- \\ \hline
    \end{tabular}
}
\vspace{-10pt}

\end{table*}

\textbf{Heterogeneous Graph Neural Network.}
HGNN follows a neighborhood aggregation scheme and a semantic fusion scheme, where the final representation of each vertex is computed by recursively aggregating the feature vectors of its neighbor vertices in each semantic graph and fusing the aggregated results across all semantic graphs, as shown in Fig. \ref{fig:HGNN}.
For example, HAN~\cite{HAN} aggregates structural information using the neighbor attention in each semantic graph and then fuses outputs from different semantic graphs using the semantic attention for each vertex.

To capture both the structural information and semantic information in HetGs, most prevalent HGNN models usually contain four major execution stages as shown in Fig. \ref{fig:HGNN}.
\blackcircledempty{1} Semantic Graph Build: The SGB stage builds semantic graphs for the following stages by partitioning the original HetG into a set of semantic graphs based on relations or predefined metapaths.
\blackcircledempty{2} Feature Projection: In the FP stage, the feature vector of each vertex is transformed to a new one using a MLP within each semantic graph.
\blackcircledempty{3} Neighbor Aggregation: The NA stage utilizes an attention mechanism~\cite{GAT} to perform a weighted sum aggregation of features from neighbors within each semantic graph.
\blackcircledempty{4} Semantic Fusion: The SF stage fuses the semantic information obtained from all semantic graphs with an attention mechanism~\cite{GAT}, aiming to combine the results of the NA stage across different semantic graphs for each vertex.
Algorithm \ref{alg:Original_Programming_Model} shows the programming model for HGNNs. 
Table~\ref{tab:HGNN_general_framework} presents the computation corresponding to each stage of four representative HGNN models.

\section{motivation}\label{sec:motivation}




This section explains the motivation to design the accelerator for HGNNs via characterization. 

\subsection{Characterization of HGNNs on GPU} \label{sec:characterization}
We conduct a quantitative characterization for HGNNs on an NVIDIA GPU T4 using two prevalent HGNN models. We focus on the last three stages for the inference acceleration on transductive learning and omit SGB stage since it is executed in CPU in the training phase. The experimental setup is shown in Section~\ref{sec:evaluation_methodology}. Note that the processing of total vertices is batched into a coarse-grained matrix operation for efficient execution on GPUs.

\begin{figure}[!t] 
	\centering
	\includegraphics[width=0.45\textwidth]{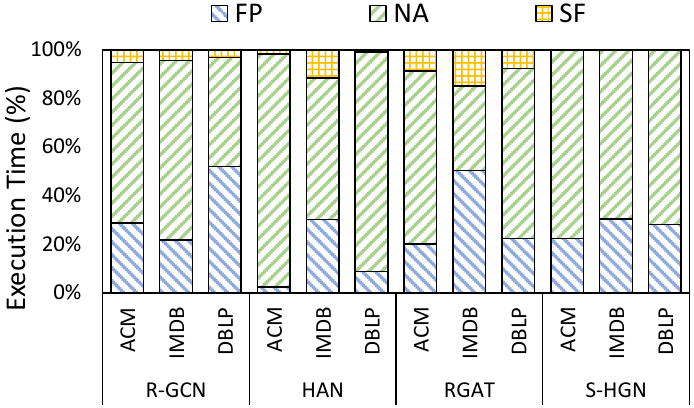}
	\vspace{-10pt}
	\caption{Execution time breakdown.} 
	\label{fig:execution_breakdown}
	\vspace{-5pt}
\end{figure}

\textbf{Execution Time Breakdown.} 
Fig.~\ref{fig:execution_breakdown} shows the breakdown of the execution time of the inference phase. 
The FP, NA, and SF stages take 26.5\%, 68.7\%, and 4.8\% execution time averaging across different models and datasets, respectively. For S-HGN, since the SF stage is replaced by the edge embedding, there is no explicit SF stage in the execution.
From above, the NA stage dominates the execution time of HGNNs. The reason is that, for each semantic graph, neighboring feature vectors of each vertex in the corresponding semantic graph need to be aggregated, which is time-consuming.

\begin{figure}[!t] 
	\centering
	\vspace{0pt}
	\includegraphics[width=0.45\textwidth]{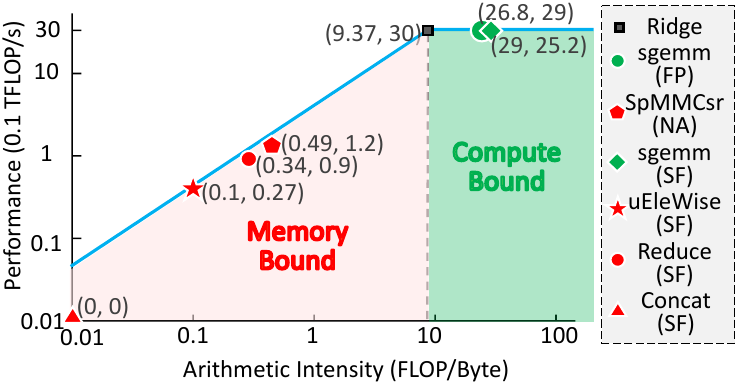}
	\vspace{-10pt}
	\caption{CUDA kernels in the floating-point operation Roofline of GPU T4 on HAN model with DBLP dataset.}
	\label{fig:roofline}
	\vspace{0pt}
\end{figure}

\textbf{Execution Bound.}
Different stages exhibit different execution bounds, leading to low utilization of various hardware components, as demonstrated in Fig.~\ref{fig:roofline} and Table~\ref{tab:kernel_profiling_details}.

$\bullet$  \textit{The FP stage is dominated by the execution of dense-dense matrix multiplication, primarily facing compute bound.} 
The CUDA kernels in the FP stage generally exhibit compute bound due to their high compute-to-memory-access ratios, such as the sgemm (dense-dense matrix multiplication) kernel. 
For example, the sgemm kernel in the HAN model on the DBLP dataset costs over 97.4\% execution time of the FP stage and achieves 95.9\% peak performance while the DRAM bandwidth is underutilized with only 33.6\%. 
The arithmetic intensity of this kernel is 26.8 FLOP/Byte and larger than the one in the ridge of Roofline (see Fig.~\ref{fig:roofline}), which reveals that the FP stage faces compute bound.

$\bullet$  \textit{The NA stage is dominated by the execution of graph-topology-dependent operations, primarily exhibiting memory bound and irregular memory access pattern.} 
The CUDA kernels in the NA stage perform compute operations dependent on the irregular topology of graphs, generally showing memory bound caused by the irregular memory accesses to neighboring feature vectors, such as the SpMMCsr (SpMM, sparse-dense matrix multiplication). 
Taking the HAN model on the DBLP dataset as an example, the SpMMCsr kernel aggregates neighboring feature vectors into a single vector for each vertex according to the irregular neighbor connection. 
This kernel achieves high DRAM bandwidth utilization (74.3\%) with a low L2 Cache hit rate (31.4\%) due to the intensive irregular memory accesses. 
Nevertheless, this kernel exhibits low arithmetic intensity (0.49 FLOP/Byte) and low achieved peak performance (3.9\%). This reveals that many computing resources in the NA stage are underutilized.

$\bullet$  \textit{The SF stage is dominated by the execution of dense-dense matrix multiplication, element-wise operation, and data rearrangement operation, primarily facing compute bound first and then memory bound.}
In this stage, the sgemm kernel first calculates attention weights for each resulting feature vector of each semantic graph from the NA stage, and then the uEleWise (unrolled\_elementwise\_kernel), Concat (CatArrayBatchedCope) and Reduce kernels aggregate these feature vectors into single one for each vertex with attention weights. 
The sgemm kernel still exhibits compute-bound as the high peak performance (84.2\%).
On the contrary, the uEleWise, Reduce, and Concat kernels show memory bound, which achieve over 80\% DRAM bandwidth utilization with minor peak performance.
As a result, the SF stage faces low DRAM bandwidth utilization first and then low compute utilization as well as low L2 Cache hit rate. 

\begin{table}[!t]
    \caption{Profiling results of major CUDA kernels on HAN model with DBLP dataset.} \label{tab:kernel_profiling_details}
    \centering
    \setlength\tabcolsep{3pt}%
	\renewcommand\arraystretch{0.8}
    \resizebox{0.49\textwidth}{!}{
\centering

\begin{tabular}{ccccc}
\toprule
  \begin{tabular}[c]{@{}c@{}} \textbf{Kernel} \\ \textbf{Name}  \end{tabular}  & 
  \begin{tabular}[c]{@{}c@{}} \textbf{Time}   \\ \textbf{(\%)} \end{tabular}     & 
  \begin{tabular}[c]{@{}c@{}} \textbf{Achieved Peak} \\ \textbf{Performance (\%)} \end{tabular} & 
  \begin{tabular}[c]{@{}c@{}} \textbf{DRAM Bandwidth} \\ \textbf{Utilization (\%)} \end{tabular}  &
\begin{tabular}[c]{@{}c@{}} \textbf{L2 Cache} \\ \textbf{Hit  Rate (\%)} \end{tabular} \\ \midrule \midrule
 \multicolumn{5}{c}{\blackcircledempty{2} Feature Projection (Compute Bound)}    \\ \midrule 
    sgemm    & 97.4\% & 95.9\% & \textbf{33.6\%} &  82.7\%   \\ \midrule \midrule
\multicolumn{5}{c}{\blackcircledempty{3} Neighbor Aggregation (Memory Bound)}       \\ \midrule 
    SpMMCsr  & 85.9\% & \textbf{3.9\%}  & 74.3\%  & \textbf{31.4\%} \\
    \midrule \midrule
    \multicolumn{5}{c}{\blackcircledempty{4} Semantic Fusion (Compute Bound $\rightarrow$ Memory Bound)}        \\ \midrule  
    sgemm   & 47.8\% & 84.2\% & \textbf{42.4\%} & 83.3\% \\
   uEleWise  & 20\%   & \textbf{0.9\%}  & 82.4\% & \textbf{50.0\%} \\
    Reduce   & 11\%   & \textbf{3.1\%}  & 88.3\%  & \textbf{25.2\%}\\  
    Concat   & 17.5\% & \textbf{0\%}    & 81.6\% & \textbf{50.0\%} \\
    \bottomrule
\end{tabular}
    }
    
  	\vspace{-5pt}
\end{table}

\textbf{Differences between GNNs and HGNNs.} 
The major differences between GNNs and HGNNs on execution are:

$\bullet$  \textit{Joint Feature Projection vs. Separate Feature Projection.} The raw feature vector of each vertex in HomoGs is in the same vector space with the same dimension. Thus, the feature projection of vertices in GNNs can be performed jointly. However, the raw feature vectors of vertices of different types in HetGs are not in the same vector space and have different dimensions, requiring different feature projection parameters. Generally, HGNNs utilize a specific feature projection matrix for each vertex type or in each semantic graph.

$\bullet$  \textit{Aggregation vs. Aggregation+Fusion.} GNNs only aggregate once for the neighbor aggregation on a single type of relation. HGNNs aggregate features from neighbors in each semantic graph generated according to corresponding semantics (relations or metapaths), and then fuse intermediate results of each semantic graph for each vertex.


\textbf{These differences introduce high-degree inter-semantic-graph parallelism and data reusability, exposing opportunities for HGNN acceleration.}
The inter-semantic-graph parallelism in the NA stage points to that different semantic graphs can be processed in parallel. The inter-semantic-graph data reusability derives from that the intermediate results of the FP stage can be reused across the processing of different semantic graphs. Furthermore, more semantic graphs can capture more semantics in deep, which helps improve the prediction accuracy of HGNNs~\cite{SeHGNN}. It follows that these parallelism and data reusability becomes higher degree as the number of metapaths increases. 

\subsection{The Need for an HGNN Accelerator}
\label{sec:the_need_for_HGNN_accelerator}

Given the above characterizations, we explain the motivation to design an HGNN accelerator.

\textbf{Limitations of GNN Accelerators.} 
GNN accelerators tailored to GNNs gain significant speedup and energy savings compared to GPUs~\cite{HyGCN,awbgcn,ReGNN}. Whereas, they lack the HGNN-oriented design and optimization to accelerate HGNNs. 
First, they have proposed stage-fusion optimizations to accelerate GNNs, but they are ill-suited to HGNNs. This is because the execution semantics and execution patterns of the sequential stages in HGNNs differ from those in GNNs, resulting in different fusion methodologies of compute, memory access, and dataflow.
Second, they miss elaborate optimizations to exploit the high-degree inter-semantic-graph parallelism and data reusability in the processing of HGNNs.
Third, GNNs and HGNNs have great differences in execution patterns. GNN accelerators lack the control unit to schedule the whole execution for HGNNs.

%


\textbf{Inefficiencies of GPUs.}
GPUs are inherently tailored to compute-intensive workloads such as dense-dense matrix multiplication~\cite{gpu_drawback}. GPUs cannot efficiently handle irregular memory accesses caused by the graph-topology-dependent execution pattern in the NA stage~\cite{HyGCN,Graphicionado,Tesseract,GraphR,GraphDynS,understand_GCN,understand_HGNN}, suffering from low efficiency. For example, the SpMMCsr kernel in the NA stage exhibits high DRAM bandwidth utilization with a low L2 Cache hit rate (see Section \ref{sec:characterization}). This reveals that frequent replacements of feature vectors have occurred, which introduce many redundant DRAM accesses. 
In addition, existing software frameworks usually execute stages with different execution bounds in serial to leverage hardware-optimized coarse-grained operations on GPUs~\cite{PyG,DGL} (e.g., SpMM), causing low compute utilization and DRAM bandwidth utilization.

\textbf{Design Requirements.}  
Given the characteristics of HGNNs, we present the architecture design requirements to perform HGNNs with high performance.
First, to improve the utilization of various hardware components, a novel stage-fusion methodology is required to fuse stages that exhibit different execution bounds.
We propose a stage-fusion programming model along with a customized hardware datapath to enable seamless pipelined execution without any stalls.
Second, the high-degree inter-semantic-graph parallelism and data reusability can be exploited to improve performance and efficiency. It is necessary to design specific computing, memory access, and control units to exploit them. 
Meanwhile, workload-aware scheduling is proposed to ensure workload balance among pipelines without increasing the latency, and similarity-aware execution scheduling is employed to remove the redundancy compute.


\section{Architecture Design}
This section presents HiHGNN, a high-performance HGNN accelerator designed to exploit both the high-degree intra- and inter-semantic-graph parallelism and data reusability. We begin by introducing a stage-fusion programming model and a novel hardware datapath coordinating with it in Section \ref{sec:stage_fusion} that effectively enables the fusion and pipeline of stages with different execution bounds. Next, we design an independency-aware parallel execution for the inter-semantic-parallelism exploitation, which involves the scale-up and workload balancing optimizations in Section \ref{sec:parallel_exploitation}. Finally, we propose a similarity-aware execution scheduling to maximize the reuse of intermediate results across the processing of semantic graphs in Section \ref{sec:data_reusability}.

\begin{algorithm}[!t]
    \SetAlgoLined
    \caption{\textbf{Stage-fusion Programming Model}}
    \footnotesize
    \label{alg:Programming Model}
    \textbf{Input:} semantic graphs of G = \{$G^{\mathcal{P}_1}$, $G^{\mathcal{P}_2}$ ... $G^{\mathcal{P}_n}$\};\\
    \textbf{Output:} the embedding $h_v$ of each vertex $v$;\\
    \textbf{Initial:} all task lists are empty. 
    
    \For{semantic graph $G^{\mathcal{P}}$ $\in$ G}{
        NA\_task\_list=$E^{\mathcal{P}}$;

            \centerline{{\underline{\color{blue} $\triangleleft \quad \textbf{FP Stage}$}}}
                    \For{each vertex v in FP\_task\_list}{
                        $h_v'$=Feature\_Projection($W^{c_v}$, $h_v$)\;
                        $\theta_v^{\mathcal{P}}$=Compute\_Coefficient(${a_{\mathcal{P}}^T}$, $h_v'$);
                    }
            \centerline{{\underline{\color{blue} $\triangleleft \quad \textbf{NA Stage}$}}}
                    \For{each edge $e_{u,v}$ in NA\_task\_list}{
                        \eIf{$x_v$ and $x_u$ is projected}{
                            $z_v^{\mathcal{P}}$$\gets$Aggregate($\theta_u^{\mathcal{P}}$, $\theta_v^{\mathcal{P}}$, $h_u'$)\;
                            \If{all neighbors of $v$ have finished NA}{
                                Send\_to\_LSF\_task\_list($v$);
                            }
                        }
                        {
                            Send\_to\_FP\_task\_list($u$,$v$);
                        }
                    }
            \centerline{{\underline{\color{blue} $\triangleleft \quad \textbf{Local SF Stage}$}}}
                \For{each vertex v in LSF\_task\_list}{
                    $w_{\mathcal{P}}$$\gets$Compute\_and\_Accumulate($q^{T}$, $z_v^{\mathcal{P}}$)\;
                    \If{all vertex in $G^{\mathcal{P}}$ have finished LSF}{
                        Send\_to\_GSF\_task\_list($w_{\mathcal{P}}$);
                    }
                }
            \centerline{{\underline{\color{blue} $\triangleleft \quad \textbf{Global SF Stage}$}}}
                \For{each $w_{\mathcal{P}}$ in GSF\_task\_list}{
                    $\beta_G$$\gets$Compute\_and\_Accumulate($w_{\mathcal{P}}$)\;
                    \For{vertex v $\in$ $V^{\mathcal{P}}$}{
                        $z_v$$\gets$Aggregate($w_{\mathcal{P}}$, $z_v^{\mathcal{P}}$)
                    }
                }
    }
    \centerline{{\underline{\color{blue} $\triangleleft \quad \textbf{Final Stage}$}}}
    \For{vertex v $\in$ $V$}{
        {$h_v$}{$\gets$}EW-DIV($z_v$,\,$\beta_G$)\;
    }
\end{algorithm}

\subsection{Bound-aware Stage Fusion}\label{sec:stage_fusion}
As mentioned in Section \ref{sec:characterization}, different stages of HGNN exhibit diverse execution bounds, leading to unbalanced utilization across different hardware components and limited performance. Although previous efforts~\cite{DNNBuilder,HyGCN, GRIP, GCoD} have employed stage-fusion techniques to achieve better parallelism in traditional workloads, they lack awareness of execution bounds and fail to match the workflow of HGNNs. This inspires us to propose a bound-aware stage fusion methodology for HGNN acceleration, efficiently improving hardware utilization and exploiting inter-stage parallelism.

\subsubsection{Stage-fusion Programming Model} \label{sec:programming_model}
To fully utilize hardware resources and implement pipeline execution, we propose a novel programming model in which the original execution stages are decomposed and reorganized, allowing the fusion and pipeline of execution for stages with different execution bounds. This model is illustrated in Algorithm~\ref{alg:Programming Model}. Note that all the functions included are user-definable, maintaining good programming flexibility and adaptability for various HGNN models. All stages except for the final stage can be executed in parallel as long as their corresponding task lists are non-empty. Overall, this differs significantly from the sequential execution of stages shown in Algorithm~\ref{alg:Original_Programming_Model}.

We have made the following modifications and optimizations compared to the original programming model.
\blackcircledempty{1}
To achieve stage fusion, we break the barrier that separates the NA stage from the FP stage in the original programming model.
The original NA stage is split into two steps, i.e., the computation of attention coefficients (lines 8) and the rest of the process (line 12).
Then we integrate the former step with the FP stage, enabling directly forwarding the projected features for the computation of attention coefficients without waiting for all vertex features to be projected.
\blackcircledempty{2}
We remove the barrier between the NA and SF stages in most HGNNs.
The barrier exists because the HGNNs require all results from different semantic graphs to generate the semantic attention importance in the NA stage.
So, we decompose the SF stage into two stages, namely the Local SF (LSF) stage and the Global SF (GSF) stage. For a given semantic graph $G^{\mathcal{P}}$, the former mainly involves the intermediate computation of the semantic attention importance (line 21), which is executed once after all the neighbors of a target vertex in $G^{\mathcal{P}}$ are aggregated. 
While the latter generates the ultimate semantic attention importance of $G^{\mathcal{P}}$ (line 27) and performs the semantic aggregation of all the target vertices (lines 28-30). It is executed once for each semantic graph when all the target vertices have accomplished their neighbor aggregation processes.
In this way, the LSF stage can be fused into the execution of the NA stage (lines 13-15).

Noticed that in the stage-fusion programming model, we implement the fine-grained parallelism between the FP stage and the NA stage. At this point, if the FP stage is still being used for driving like in the original programming model, then every time a vertex is projected in the FP stage, the NA stage has to search for its neighbors in the edge list or adjacency matrix, which triggers a large number of random accesses and degrades the performance. To reduce the random accesses, we put all edges in the NA\_task\_list, which eliminates the random access for neighbor search. In this way, although the access to the feature becomes random, the feature itself has a larger dimension, thus the impact on overall bandwidth utilization is relatively small.

\begin{figure}[!t] 
	\centering
	\includegraphics[width=0.48\textwidth]{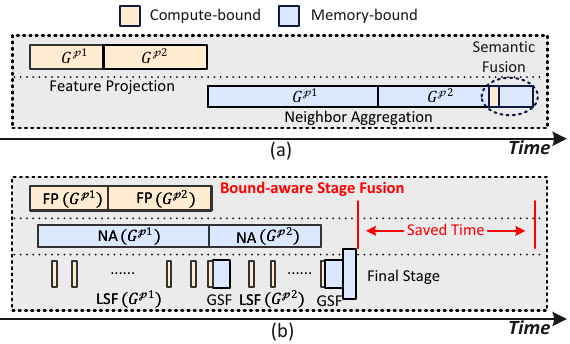}
	\vspace{-10pt}
	\caption{Execution flow of HGNN programming model: (a) Traditional staged programming model; (b) Bound-aware stage-fusion programming model.}
	\label{fig:programming_model}
\end{figure}

\begin{figure}[!b] 
	\centering
	\includegraphics[width=0.48\textwidth]{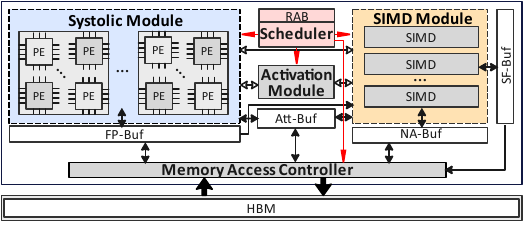}
	\vspace{-10pt}
	\caption{Basic design of HiHGNN architecture.}
	\label{fig:hardware_components}
\end{figure}

Fig.~\ref{fig:programming_model} illustrates the difference in the execution flow between the traditional one (Fig.~\ref{fig:programming_model} (a)) and our programming model (Fig.~\ref{fig:programming_model} (b)). By decomposing and fusing the execution flow, we can combine the execution of the compute-bound FP stage and the memory-bound NA stage. Besides, we can also fuse the compute-bound LSF stage with its preceding memory-bound NA stage to improve the utilization of compute and bandwidth resources simultaneously. Additionally, we pipeline the execution of each stage, enabling the different stages to run in parallel, thus resulting in performance improvement.



\subsubsection{Implementation of Hardware Datapath}\label{sec:Hardware_datapath}
In conjunction with the proposed programming model, we implement several hardware components to constitute to hardware datapath in microarchitecture. We employ a hybrid architecture to minimize the design complexity of hardware components, as in work~\cite{TPU,HyGCN,SGCN,ReGNN}. This enables us to facilitate data path optimization and leverage parallelism as well as data reusability more effectively.



\textbf{Hardware Components.} 
Throughout the execution of the HGNN models, computational workloads are mainly occupied by matrix-vector multiplication (MVM) and element-wise (EW) operations. To boost their execution efficiency, we design a dedicated module for each of them as depicted in Fig.~\ref{fig:hardware_components}. First, a flexible \textit{Systolic Module} is established for the MVM operations, which is based on the well-known systolic array design \cite{TPU}. Two execution modes, one for fine-grained matrix multiplication and the other for larger-scale matrix operation, are employed as suggested in work \cite{HyGCN}. Second, EW operations running through the whole processing of HGNNs are performed by the \textit{SIMD Module}. In addition, the \textit{Activation Module} is built to perform non-linear functions like \textit{LeakyRelu}, \textit{Elu}, and \textit{Softmax}. To coordinate workloads across various hardware components, we introduce a centralized \textit{Scheduler} equipped with a redundancy-aware bitmap (RAB) which is built by a set of two-port SRAM banks to support the reuse of intermediate results. The RAB will be further explained in Section~\ref{sec:basic_implementation_of_data_reuse}.

The high-bandwidth memory (HBM) stores information about the original semantic graphs, mainly including the adjacent information in the compressed sparse column (CSC) format and raw features stored continuously according to vertex categories. \textit{Feature Projection Buffer} (FP-Buf) and \textit{Neighbor Aggregation Buffer} (NA-Buf) aim to cache the projected features and the intermediate neighbor aggregated features respectively for data reuse. Similar to NA-Buf, \textit{Semantic Fusion Buffer} (SF-Buf) is used to hold the features that are aggregated from multiple semantic graphs for vertices. During the computation process of the attention mechanism, a series of parameters as well as intermediate results, are cached in the \textit{Attention Buffer} (Att-Buf). The \textit{Memory Access Controller} is used to schedule the data interactions between the on-chip buffers and HBM.


\textbf{Hardware Datapath.} The hardware datapath of our proposed stage-fusion programming model is built from the fundamental hardware components we stated above and represents the dataflow during execution. 

The main challenge in building the hardware datapath is to remove the pipeline stall. 
On one hand, the data access conflicts caused by fine-grained parallelism with edge-centric granularity increase the access latency, and on the other hand, the accumulation of \textit{Softmax} functions on the denominator also leads to pipeline stalls. For the former we can utilize multi-port SRAM to realize on-chip buffers, while for the latter we need to further decompose the execution flow of \textit{Softmax}. Take the weighted aggregation of NA stage as an example, the decomposition is as follows: 

%

\begin{equation*}
\setlength{\abovedisplayskip}{2pt}
\setlength{\belowdisplayskip}{2pt}
    \begin{aligned}  \label{eq:softmax decompose}
        \displaystyle \sum_{u\in \mathcal{N}_v^{\mathcal{P}}} \alpha_{u,v}^{\mathcal{P}} \cdot h_u^{'}=\frac{\displaystyle \sum_{u\in \mathcal{N}_v^{\mathcal{P}}} exp(\theta_{u,v}^{\mathcal{P}}) \cdot h_u^{'}}{\displaystyle \sum_{k\in \mathcal{N}_v^{\mathcal{P}}} exp(\theta_{k,v}^{\mathcal{P}})}
    \end{aligned}
\end{equation*}

\begin{figure}[!htbp] 
	\centering
	\includegraphics[width=0.48\textwidth]{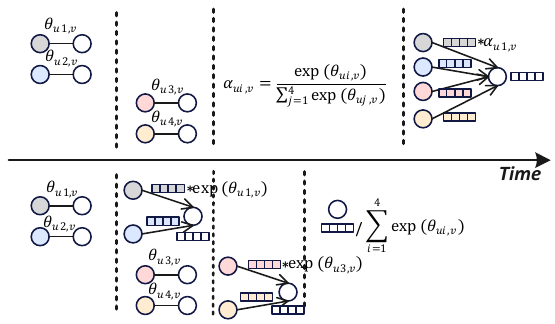}
	\vspace{-10pt}
	\caption{Optimized execution flow of Softmax function.}
	\label{fig:softmax_decompose}
\end{figure}

The weighted aggregation in the GSF stage also employs a similar approach. An example is illustrated in Fig.~\ref{fig:softmax_decompose}. Once the numerator is computed, it can be immediately used for aggregation and accumulated onto the denominator, avoiding pipeline stalls caused by waiting for the computation of \textit{Softmax} denominator.

\begin{figure*}[!t] 
	\centering
	\includegraphics[width=1\textwidth]{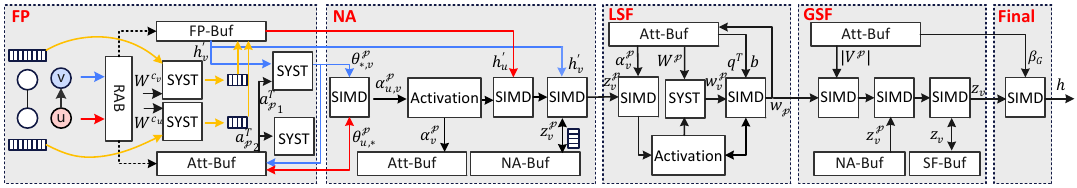}
	\vspace{-20pt}
	\caption{Hardware datapath for the stage-fusion programming model.}
	\label{fig:datapath}
	\vspace{-10pt}
\end{figure*}

The resulting datapath is depicted in Fig.~\ref{fig:datapath}. For each edge $e_{u,v}$, the raw features of the source vertex $u$ and the target vertex $v$ are projected into the same vector space through a linear transformation function performed by the \textit{Systolic Module} (indicated by SYST in Fig.~\ref{fig:datapath}). A bitmap-based data reuse mechanism is employed here to remove computation redundancy. Since $h_u'$ and $h_v'$ are globally reusable while $\theta_{u,*}^{\mathcal{P}}$ and $\theta_{*,v}^{\mathcal{P}}$ ($\theta^{\mathcal{P}}_{u, *}$ for source vertex and $\theta^{\mathcal{P}}_{*, v}$ for target vertex) are only reusable within the same semantic graph, a total of three cases arise for the reuse mechanism, as indicated by the different colored lines in Fig.~\ref{fig:datapath}. More details are given in Section~\ref{sec:basic_implementation_of_data_reuse}. 

Unlike the traditional HGNN execution process, after obtaining $h'_{u}$ and $h'_{v}$, instead of writing them back and keeping them until aggregation in the NA stage, they are immediately sent to the \textit{Systolic Module} for the computation of $\theta_{u,*}^{\mathcal{P}}$ and $\theta_{*,v}^{\mathcal{P}}$ in NA stage, and the weighted aggregation operation from $u$ to $v$ is completed by the \textit{SIMD Module} afterward. 

For each target vertex $v$ in $G^{\mathcal{P}}$, once all the features of neighbors have been aggregated by the \textit{SIMD Module}, a partial computation of the semantic importance $w_{\mathcal{P}}$ contributed by $v$ denoted as $w_v^{\mathcal{P}}$ is carried out using the \textit{Systolic Module} instantly. Meanwhile, the value is accumulated in the ultimate result of $w_{\mathcal{P}}$. When all vertices in $G^{\mathcal{P}}$ finish the FP, NA, and LSF stages, the GSF stage first computes the ultimate semantic importance $w_{\mathcal{P}}$ of $G^{\mathcal{P}}$ by dividing it with the total number of vertices $V^{\mathcal{P}}$ and then aggregates the semantic features of all vertices in $G^{\mathcal{P}}$, which is similar to NA and is performed on \textit{SIMD Module}. Following the completion of all the aforementioned stages, the final embeddings $h$ are produced by dividing target vertices' features by an accumulated global weight $\beta_G$.



\begin{figure}[!b] 
    \vspace{-5pt}
	\centering
	\includegraphics[width=0.48\textwidth]{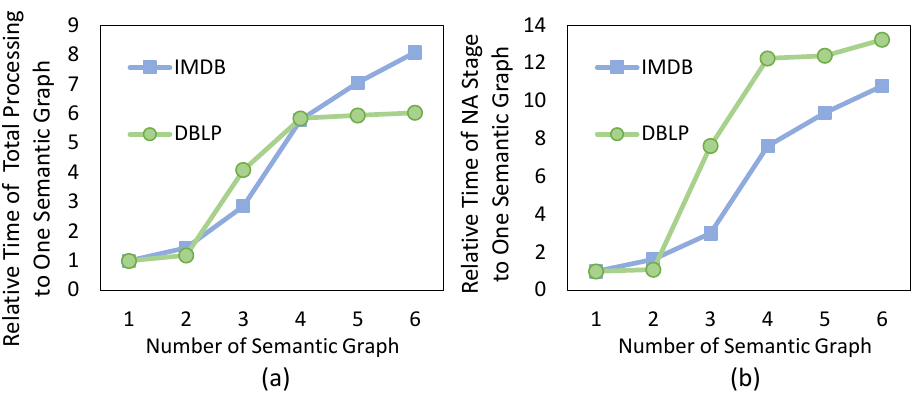}
	\vspace{-10pt}
	\caption{Relative time of execution to one semantic graph: (a) Time of total execution; (b) Time of NA stage.}
	\label{fig:relative_time}
	\vspace{-5pt}
\end{figure}

\subsection{Parallelism Exploitation: Independency-aware Parallel Execution}\label{sec:parallel_exploitation}
In this section, we propose the independency-aware parallel execution to exploit the inter-semantic-graph parallelism by leveraging the independency among semantic graphs.

\subsubsection{Scale-up Optimization for Basic Architecture} \label{sec:multi-lane}
To dig into the parallelism across semantic graphs, we provide the scale-up optimization for the above architecture. 

We first give the motivation to exploit inter-semantic-graph parallelism. First, the processing of each semantic graph in the NA stage is independent.
Second, the NA stage dominates the total execution time of HGNNs, as demonstrated in Section~\ref{sec:motivation}.
Third, the execution time of total execution and the NA stage both increase as the number of semantic graphs increases, as demonstrated in Fig.~\ref{fig:relative_time} (a) and (b). 
Therefore, it is vital to exploit the parallelism across semantic graphs.

\begin{figure*}[!htbp] 
	\centering
	\includegraphics[width=1.0\textwidth]{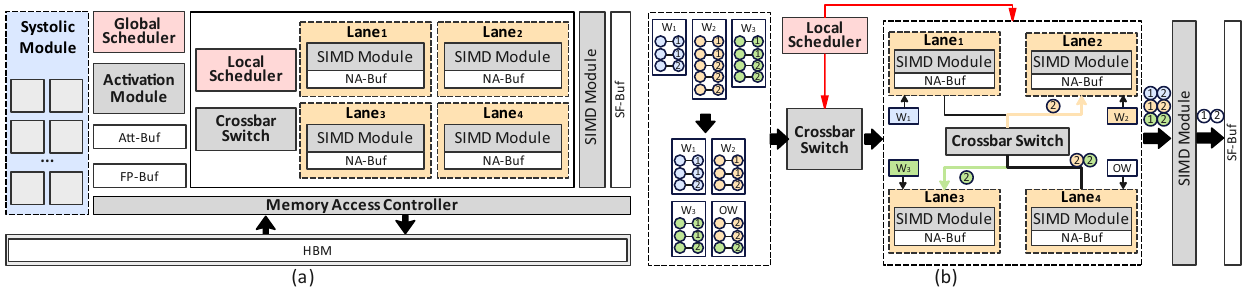}
	\vspace{-20pt}
	\caption{Independency-aware parallel execution: (a) Scale-up optimization for basic architecture; (b) Workload balance by workload-aware scheduling.}
	\label{fig:multi-lane}
\end{figure*}

To this end, a scale-up optimization for the basic architecture as shown in Fig.~\ref{fig:multi-lane} (a) is presented. 
The basis for scaling up is the extension of the lanes and each lane independently processes a semantic graph.
For simplicity, we refer to the basic design of HiHGNN as a single-lane architecture, while the scale-up design of HiHGNN is called a multi-lane architecture.
In the multi-lane architecture, the hardware components are generally the same as those in the single-lane one. The \textit{Systolic Module} is mainly responsible for MVM operations, while the \textit{SIMD Module} handles EW operations. To schedule the execution, we replace the original \textit{Scheduler} with \textit{Global Scheduler} which schedules the lane group with other modules. A scheduler named \textit{Local Scheduler} is used to handle the processing within the lane group.
An additional \textit{SIMD Module} outside to execute the GSF stage after synchronizing the results of different lanes.
Besides, to allocate the workload to each lane and achieve dynamic scheduling across them, we leverage crossbar switches with the number of ports being consistent with the number of lanes for data transmission, which is illustrated in detail in the next section.

\subsubsection{Workload Balance across Hardware Datapaths} \label{sec:workload_balance}
To make full use of compute units, we propose workload-aware scheduling to balance the workload across different lanes. 

With the scale-up design, the multi-lane architecture can leverage both stage-level parallelism and semantic-graph-level parallelism.
However, the workload imbalance across different semantic graphs is severe in real-world datasets. 
For example, the DBLP dataset consists of three semantic graphs, each containing 4057 vertices, but with vastly different numbers of edges, i.e., 7043571, 5000496, and 11113, respectively.
When the three workloads are assigned directly to the $Lane_1$, $Lane_2$, and $Lane_3$. $Lane_1$ and $Lane_2$ become overloaded, while $Lane_3$ is underloaded. 

Additionally, to avoid introducing extra overhead during the execution process, we require a low-latency and stable algorithm to balance the workloads.
Fortunately, unlike the situation in the previous work~\cite{workloadbalance}, there are no computational dependencies between different workloads. This advantageous characteristic allows us to adopt a low-cost strategy to achieve workload balance among lanes.

To fully utilize the compute resources, we propose workload-aware scheduling illustrated in Fig.~\ref{fig:multi-lane} (b) on a four-lane architecture.
We denote the initial workloads received by the global scheduler from three semantic graphs during a specific period as $W_1$, $W_2$, $W_3$, respectively, and the four lanes as $Lane_1$, $Lane_2$, $Lane_3$ and $Lane_4$. The allocation threshold for each lane is set to the maximum number of edges that the lane can process once. In this example, the threshold is set to three.
During the runtime, the \textit{Local Scheduler} first identifies each task list that exceeds the threshold and pushes the excess parts into a task list \textit{Overflow Workload} ($OW$) to prevent any blocking in the execution. 
Once none of the task lists exceeds the threshold, the \textit{Local Scheduler} assigns the task lists to their corresponding lanes, such as $W_1$ to $Lane_1$, except for the $OW$.
Finally, the \textit{Local Scheduler} assigns the workloads in the $OW$ to the lanes that have not reached the threshold. 
This scheduling approach ensures that there is no blocking in each lane and maximizes the utilization of the compute units.

For synchronization of intermediate results, the \textit{Local Scheduler} records the aggregation status of each vertex. When a vertex finishes NA, all lanes that have the partial aggregation results of this vertex send these results to the lane it is originally assigned to in the workloads dispatch.
As an example, in this case, the vertices executed on $L_4$ are transferred to $L_2$ and $L_3$ for the LSF stage. When all vertices within a semantic graph have been finished the LSF stage, the results are sent to an outside \textit{SIMD Module} for the GSF stage and then stored in SF-Buf. 

\subsection{Data Reusability Exploitation: Similarity-aware Execution Scheduling} \label{sec:data_reusability}
Due to the fine-grained execution flow, it inevitably introduces a significant amount of redundant computations and DRAM accesses. 
To address this issue and eliminate redundant computations and DRAM accesses within and across the processing of semantic graphs, we initially employ a basic data-reuse technique.
Building on this foundation, we further propose a similarity-aware execution scheduling to exploit the inter-semantic-graph data reusability.

\subsubsection{Basic Mechanism of Data Reuse} \label{sec:basic_implementation_of_data_reuse}
To harvest the reusability of intermediate results both within and across semantic graphs, we implement a basic mechanism for data reuse using a bitmap.

We first give two findings to show the data-reuse opportunity. 
i) For the prevalent HGNN models, 
the projected features can be reused within and across semantic graphs in FP stage due to the vertex-type-specific projection. ii) As illustrated in Algorithm~\ref{alg:Programming Model}, the attention importance used for the aggregation from the source vertex $u$ to the target vertex $v$ in the NA stage can be generated from $\theta^{\mathcal{P}}_{u, *}$ and $\theta^{\mathcal{P}}_{*, v}$, which are derived from $h'_u$ and $h'_v$ directly. 
These coefficients can be reused for all edges, avoiding recomputations for each edge.

\begin{figure*}[!t] 
    \small
    \centering
    \includegraphics[width=1.0\textwidth]{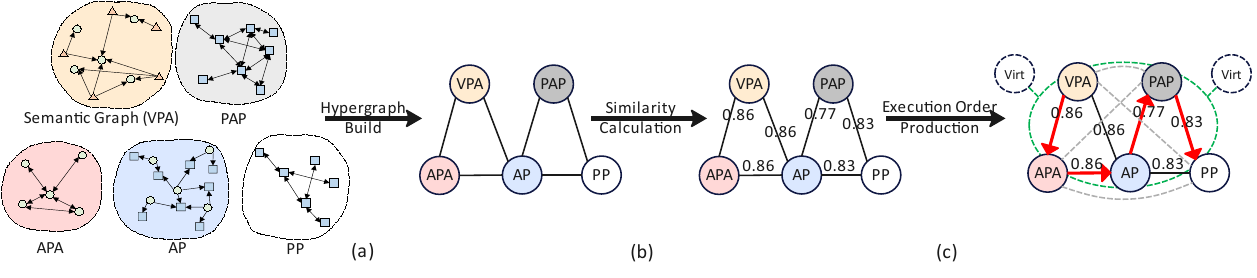}
    \vspace{-20pt}
    \caption{Similarity-aware execution scheduling: (a) Hypergraph build; (b) Similarity calculation; (c) Execution orders production by shortest Hamilton path algorithm.}
    \label{fig:hypergraph}
    \vspace{-10pt}
\end{figure*}

\begin{table}[!t]
    \centering
    \vspace{-10pt}
    \caption{The status of feature vector encoded by RAB.}
    \label{table:rab encoding}
    \resizebox{0.37\textwidth}{!}{
        \begin{tabular}{|c|ccc|}
        \hline
         \multirow{2}{*}{\diagbox{}{}}  & \multicolumn{3}{c|}{Status}                                                                         \\ \cline{2-4} 
                              & \multicolumn{1}{c|}{Projected} & \multicolumn{1}{c|}{Calculated $\theta^{\mathcal{P}}_{u, *}$} & Calculated $\theta^{\mathcal{P}}_{*, v}$ \\ \hline
        \multirow{5}{*}{Code} & \multicolumn{1}{c|}{0}         & \multicolumn{1}{c|}{0}                     & 0                     \\ \cline{2-4} 
                              & \multicolumn{1}{c|}{1}         & \multicolumn{1}{c|}{0}                     & 0                     \\ \cline{2-4} 
                              & \multicolumn{1}{c|}{1}         & \multicolumn{1}{c|}{1}                     & 0                     \\ \cline{2-4} 
                              & \multicolumn{1}{c|}{1}         & \multicolumn{1}{c|}{0}                     & 1                     \\ \cline{2-4} 
                              & \multicolumn{1}{c|}{1}         & \multicolumn{1}{c|}{1}                     & 1                     \\ \hline
        \end{tabular}
    }
    \vspace{-5pt}
\end{table}

Based on these findings, we implement a bitmap called RAB to keep track of whether the computation for a given vertex has been performed or not. 
We record both the projected features and the attention coefficients to eliminate redundant computation within and across semantic graphs. 
However, since each vertex in every semantic graph has multiple status, providing separate bitmaps for each individual case will significantly increase overhead. Thus, we integrate the RAB into the \textit{Local Scheduler} and employ the binary encoding approach to assign three bits to each vertex, which are used to record its status.
These bits are indexed by the pair (type, index). The encoding of the bitmap is shown in Table~\ref{table:rab encoding}. 
Before starting the feature projection for a vertex $v$, the \textit{Global Scheduler} first checks the bitmap for its corresponding bits. If it is filled with full zeros, the vertex's current feature vector needs to be projected. Similarly, the middle and lowest bits indicate whether the attention coefficients in that semantic graph corresponding to the vertex have been generated.


Meanwhile, a raw feature is solely accessed from DRAM for the first projection of a vertex in scenarios where RAB is utilized. Subsequently, the projected features and attention coefficients can be directly read from either DRAM or on-chip buffer for subsequent computations, thereby eliminating the need for repetitive access to raw features in DRAM.

\subsubsection{Similarity-aware Execution Scheduling}
\label{sec:SES}
To further exploit the data reusability across semantic graphs, we propose a similarity-aware execution scheduling by adjusting the execution order of semantic graphs based on the similarity between semantic graphs. All the scheduling occurs in the preprocessing, which can be included in the SGB stage.

For the prevalent HGNN models, each type of vertex is projected only once across all semantic graphs, and stored in the FP-Buf for reuse. 
Therefore, if there is a significant overlap (similarity) in vertex types between two consecutive semantic graphs, the latter semantic graph can reuse the projected feature left in the FP-Buf by the former semantic graph. This can help avoid duplicate DRAM accesses and improve performance.
For instance, considering that the heterogeneous graph includes semantic graphs \{AP, PT, PP, APA, AVA\}, one of the best scheduling orders is APA-AVA-AP-PP-PT. In this way, the semantic graph APA loads the feature of vertex A, which can be left in the FP buffer and reused by the subsequent semantic graphs such as AVA and AP; after the semantic graph AP is executed, the feature of vertex P replaces the feature of vertex A in the FP buffer and can be reused by the following semantic graphs PP and PT.

To identify the reusability of data between semantic graphs, we represent each semantic graph as a vertex in a hypergraph, as shown in Fig.~\ref{fig:hypergraph} (a). In the hypergraph, each edge connects two semantic graphs that have at least one common type of vertex. This means that the projected features of the common vertices can be reused, promoting computational efficiency. To measure the similarity between two connected semantic graphs, we assign a weight to the edge that reflects their similarity as shown in Fig.~\ref{fig:hypergraph} (b). The weight is calculated using the following formula: $w_e=1 - \eta_e/\sum_{i \in E} \eta_i$, where \textit{E} is the set of all edges in the hypergraph and $\eta_e$ is the number of common vertices in the two semantic graphs connected by the edge \textit{e}. A lower weight value of $w_e$ indicates a higher similarity between the two semantic graphs, implying that more projected features can be reused.

After the hypergraph is built, the scheduling of the execution order of semantic graphs is transformed to finding a path that starts from a specific semantic graph and passes through all other semantic graphs exactly once.
The objective is to maximize the similarity across all sequentially executed semantic graphs, thereby achieving more reuse of intermediate results. To accomplish this, we utilize the \textit{Shortest Hamilton Path} algorithm on the hypergraph to determine a specific order sequence of semantic graphs for execution.

There are still two challenges need to be addressed. First, not all hypergraphs created by semantic graphs are \textit{Hamiltonian graphs}, which means that \textit{Hamilton paths} are not always found. Second, since there is no inherent execution priority among semantic graphs, the beginning and ending points of the \textit{Hamilton path} are random. Therefore, as shown in Fig.~\ref{fig:hypergraph} (c), we first add extra edges to the hypergraph with a weight of 1, represented by gray dashed connections, which makes hypergraph a complete graph. Second, we add two virtual vertices to the hypergraph, which are connected to all other vertices with weights of zero, represented by the green dashed connections with virtual vertices. These virtual vertices are then used as the beginning and ending points of the \textit{Hamilton path}, respectively.
Finally, the execution orders of semantic graphs are produced by the shortest Hamilton path algorithm.


\section{Experimental Methodology}\label{sec:evaluation_methodology}
\textbf{Methodology.} The performance and energy of HiHGNN are evaluated using the following tools.

\textit{Cycle-accurate Simulator.} 
For the performance evaluation, a customized cycle-accurate simulator is designed and implemented to measure execution time in the number of cycles. This simulator models the microarchitectural behavior of each hardware module of HiHGNN. In addition, a detailed cycle-accurate on-chip memory model is implemented and integrated. This is also integrated with Ramulator~\cite{ramulator}, a cycle-accurate DRAM simulator, to simulate the cycle-accurate behavior of DRAM accesses to HBM.

\textit{CAD Tools.} 
For area, power, and critical path delay measurements, we implement an RTL version of each hardware module and synthesize it. We use the Synopsys Design Compiler with the TSMC 12 $nm$ standard VT library for the synthesis and estimate the power consumption using Synopsys PrimeTime PX. The slowest module has a critical path delay of 0.83 $ns$ including setup and hold time, putting HiHGNN comfortably at the 1 GHz clock frequency. 

\textit{Memory Measurements.} 
The access latency, energy, and area of the on-chip buffers are estimated using Cacti 6.5~\cite{CACTI}. We use four different scaling factors to convert them to 12 $nm$ technology, as depicted in work~\cite{technology_scale,ozdal_energy_2016} since Cacti only supports down to 32 $nm$ technology. 
The access latency and energy of HBM1.0 are simulated by Ramulator and estimated with 7 pJ/bit as in the work~\cite{7pj}, respectively.

\begin{table}[!t]
\vspace{-5pt}
\centering
\caption{Information of HetG datasets.} \label{tab:datasets}
\renewcommand\arraystretch{1.2}
\setlength\tabcolsep{2pt}%
\resizebox{0.49\textwidth}{!}{
\begin{tabular}{ccccc}
\toprule
\textbf{Dataset} &
  \textbf{\#Vertex} &
  \textbf{\#Feature} &
  \textbf{\#Edge of Each Relation} &
  \textbf{Metapath} \\ \midrule
\multirow{4}{*}{IMDB} & movie (M): 4932 & M: 3489  & 
\  \multirow{4}{*}{\begin{tabular}[c]{@{}c@{}}AM: 14779 MA: 14779\\ KM: 23610 MK: 23610\\ DM: 4932 MD: 4932\end{tabular}} &
\multirow{4}{*}{\begin{tabular}[c]{@{}c@{}}{MDM}\\ {MAM}\\ {MKM}\end{tabular}}\\
 & director (D): 2393 & D: 3341 &     &  \\
 & actor (A): 6124    & A: 3341 &     &  \\
 & keyword (K): 7971  & K: --- &     &  \\ \midrule
\multirow{4}{*}{ACM} &
  paper (P): 3025 &
  P: 1902 &
  \multirow{4}{*}{\begin{tabular}[c]{@{}c@{}}TP: 255619 PT: 255619\\ SP: 3025 PS: 3025\\ PP: 5343 -PP: 5343 \\ AP: 9949 PA: 9949\end{tabular}} &
  \multirow{4}{*}{\begin{tabular}[c]{@{}c@{}}{PPSP}\\ {PSP}\\ {PPAP}\\ {PAP}\end{tabular}} \\
 & author (A): 5959   & A: 1902 &        &  \\
 & subject (S): 56   & S: 1902 &        &  \\
 & term (T): 1902     & T: --- &        &  \\ \midrule
\multirow{4}{*}{DBLP} & author (A): 4057 & A: 334 & 
\multirow{4}{*}{\begin{tabular}[c]{@{}c@{}}AP: 19645 PA: 19645\\ VP: 14328 PV: 14328\\ TP: 85810 PT: 85810\end{tabular}} &
\multirow{4}{*}{\begin{tabular}[c]{@{}c@{}}{APA}\\ {APTPA}\\ {APCPA}\end{tabular}} \\
 & paper (P): 14328   & P: 4231 &        &  \\
 & term (T): 7723     & T: 50   &        &  \\
 & venue (V): 20      & V: --- &        &  \\ \bottomrule
\end{tabular}
}
\scriptsize
\end{table}

\textbf{Benchmark Datasets and HGNN Models.} 
The benchmark datasets that we used are listed in Table~\ref{tab:datasets}.
The raw features of each vertex are provided by the dataset itself.
The scale of heterogeneous graph datasets differs from that of homogeneous graph datasets in terms of calculation methods. The size of homogeneous graph datasets is determined by the number of edges and vertices they exhibit, while the size of heterogeneous graph datasets is determined by the construction of semantic graphs by the metapaths. For instance, in the experiment, the three datasets include 92,760, 82,384, and 133,910 vertices, and 630,542, 6,309,372, and 24,669,366 edges, respectively.
Four popular HGNN models are used, including HAN~\cite{HAN}, R-GCN~\cite{R-GCN}, R-GAT~\cite{R-GAT}, and S-HGN~\cite{Simple-HGN}, which are widely adopted in the evaluation of algorithm community~\cite{Simple-HGN,HGL,hgnn_survey_hanjiawei}.



\textbf{Baseline Platforms.} To compare the performance of HiHGNN to the state-of-the-art work, all HGNN models are implemented using a state-of-the-art framework DGL 1.0.2 \cite{DGL} and evaluated on an NVIDIA GPU T4 and an NVIDIA GPU A100, using the NVIDIA Nsight Compute. All models are implemented with the same number of hidden units \{64\} and layers are \{1, 3, 3, 2\} for \{HAN, R-GAT, R-GCN, S-HGN\}, respectively.
In addition, we implement the above HGNN models in HiHGNN. All models are implemented with the same number of hidden units and layers used in GPUs. Table \ref{tb:platform} lists the configurations for the above implementations. 

It should be noted that utilizing our proposed optimizations in GPU platform may result in the loss of hardware-optimized benefits on GPUs, which could outweigh the potential gains. This is because: i) GPUs accelerate HGNNs using coarse-grained hardware-optimized operations (i.e., SpMM) which are sophisticatedly optimized in hardware level; ii) Our optimizations are based on decoupling coarse-grained operations into finer-grained ones.

\begin{table}[!t]
\centering
\caption{Platforms for HiHGNN and baselines.} \label{tb:platform}
\renewcommand\arraystretch{1.2}
\setlength\tabcolsep{2pt}%
\resizebox{0.48\textwidth}{!}{
\begin{tabular}{ccccll}
\toprule
 & \textbf{T4} & \textbf{A100}   & \multicolumn{3}{c}{\textbf{HiHGNN (4 Lanes)}}                                                                                                 \\ \midrule
\textbf{\begin{tabular}[c]{@{}c@{}}Peak\\ Performance\end{tabular}}  & \begin{tabular}[c]{@{}c@{}} 8.1 TFLOPS, \\  1.59 GHz \end{tabular}                 & \begin{tabular}[c]{@{}c@{}} 19.5 TFLOPS, \\  1.41 GHz \end{tabular}                & \multicolumn{3}{c}{\begin{tabular}[c]{@{}c@{}} 16.38 TOPS, \\  1.0 GHz \end{tabular}} \\ \midrule
\textbf{\begin{tabular}[c]{@{}c@{}}On-chip\\ Buffer\end{tabular}}  & \begin{tabular}[c]{@{}c@{}}L1 Cache 1.28 MB,\\ L2 Cache 4 MB \end{tabular} & \begin{tabular}[c]{@{}c@{}}L1 Cache 20 MB,\\ L2 Cache 40 MB \end{tabular} & \multicolumn{3}{c}{\begin{tabular}[c]{@{}c@{}}2.44 MB (FP-Buf),\\14.52 MB (NA-Buf),\\0.12 MB (SA-Buf),\\ 0.38 MB (Att-Buf)\end{tabular}} \\ \midrule
\textbf{\begin{tabular}[c]{@{}c@{}}Off-chip\\ Memory\end{tabular}} & \begin{tabular}[c]{@{}c@{}} 300 GB/s, \\ GDDR6\end{tabular}                                & \begin{tabular}[c]{@{}c@{}}1935 GB/s, \\ HBM2e\end{tabular}                                & \multicolumn{3}{c}{\begin{tabular}[c]{@{}c@{}}512 GB/s, \\ HBM1.0\end{tabular}}                                                                                                                            \\ \bottomrule
\end{tabular}
}
\scriptsize
\vspace{0.5pt}
\par Note: The compute units include 96 systolic arrays (each with 8$\times$8 MACs) and 128 8-way SIMD cores for each lane.
\vspace{-5pt}
\end{table}

\section{Experimental Results} \label{sec:experimental_results} 
In this section, we compare HiHGNN with the baselines and present the optimization analysis in detail. 
The last set of bars in the figures, labeled GM, shows the geometric mean across all HGNN models.

\subsection{Overall Results} \label{sec:overall_results} 

\begin{figure*}[!hptb] 
    \vspace{-5pt}
    \small
    \centering
    \includegraphics[width=0.98\textwidth]{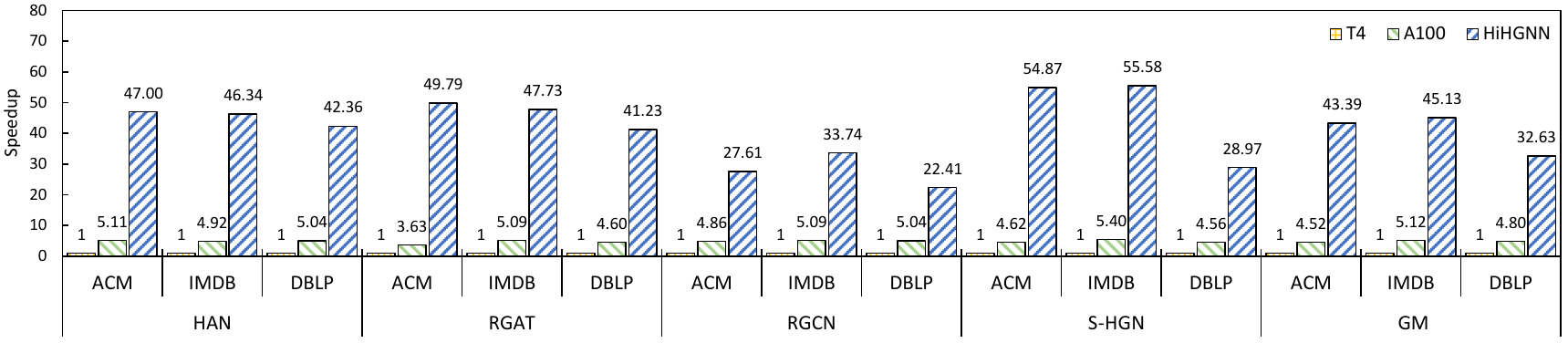}
    \vspace{-10pt}
    \caption{Speedup over GPU T4.}
    \label{fig:overall_speedup}
    \vspace{-15pt}
\end{figure*}

\textbf{Speedup.} 
Fig.~\ref{fig:overall_speedup} shows the speedup of HiHGNN to GPU T4.
HiHGNN achieves an average speedup of 40.0$\times$ and 8.3$\times$ compared to GPU T4 and A100, respectively.
The performance improvement of HiHGNN is attributed to the well-designed hardware datapath for HGNN and the efficient exploitation of inter-semantic-graph parallelism and data reusability. First, the bound-aware stage-fusion programming model and its corresponding hardware datapath help greatly improve compute utilization and bandwidth utilization across all stages.
Second, an independency-aware parallelism exploitation optimization helps leverage more hardware resources to improve performance.
Third, the similarity-aware execution scheduling significantly reduces the random accesses to DRAM, resulting in further performance improvement.
The detailed analysis of the effects of these optimizations is presented in Section \ref{sec:evaluation_optimization}.

\textbf{Compare with HyGCN-like Accelerator.}
As discussed in Section~\ref{sec:the_need_for_HGNN_accelerator}, existing HyGCN-like accelerators primarily cater to GCN models and lack support for the execution flow of HGNNs. To underscore the advantages of HiHGNN over HyGCN-like accelerators, we modify the execution flow of HiHGNN to align with that of HyGCN-like accelerators, denoting the modified version as HyHGNN. This adaptation involves na\"ive stage fusion, shifting the parallel granularity from the edge level to the stage level.
It is worth noting that the techniques introduced by HiHGNN, such as workload-aware scheduling and similarity-aware execution scheduling, are not incorporated into HyHGNN. 

\begin{figure}[!hptb] 
    \vspace{-5pt}
    \small
    \centering
    \includegraphics[width=0.48\textwidth]{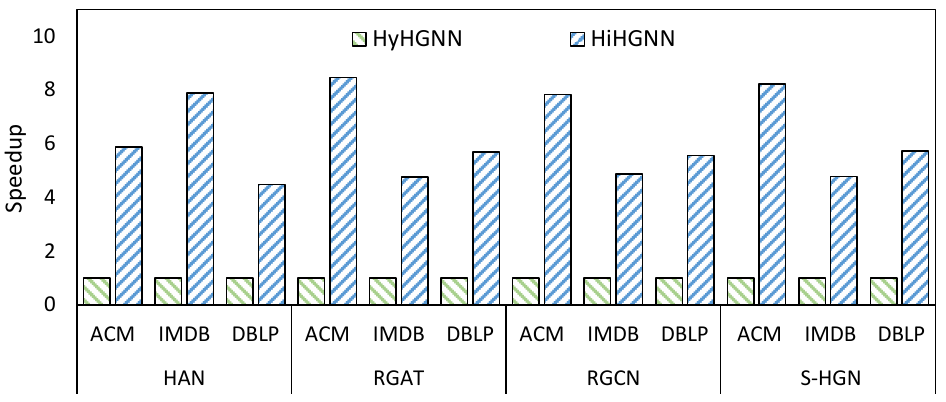}
    \vspace{-10pt}
    \caption{The advantage brought by Stage-fusion Programming Model over Na\"ive Stage Fusion.}
    \label{fig:compare_simple_stage_fusion}
    \vspace{-5pt}
\end{figure}

Fig.~\ref{fig:compare_simple_stage_fusion} manifests that the HiHGNN yields a 6.1$\times$ improvement over HyHGNN. 
This improvement primarily stems from the dedicated datapath design tailored for HGNNs. Additionally, HiHGNN's superior parallelism, facilitated by stage-fusion programming and independency-aware parallel execution, contributes to the enhanced performance. Furthermore, the efficient data reuse mechanism of HiHGNN, specifically the similarity-aware execution scheduling, plays a crucial role in driving these performance gains.

\textbf{Area and Power.} Table \ref{tb:chip_area_power} provides a detailed breakdown of the area and power of HiHGNN (except for DRAM), which are 21.36 $mm^2$ and 12.00 $W$, respectively. The buffers that include FP-Buf, NA-Buf, SF-Buf, and Att-Buf occupy 60.25\% area and consume 12.10\% power. \textit{SIMD Module} and \textit{Systolic Module} are the main computing units that produce 38.81\% and 83.61\% power, respectively. For the computation precision, we use a 32-bit integer that is enough to maintain the accuracy of HGNN inference.

\begin{table}[!t]
\vspace{-5pt}
 \caption{Characteristics of HiHGNN (TSMC 12 $nm$).} \label{tb:chip_area_power}
 \centering
 \renewcommand\arraystretch{1.0}
    \resizebox{0.46\textwidth}{!}{
\begin{tabular}{|l|l|r|r|r|r|}
\hline
\multicolumn{2}{|l|}{\begin{tabular}[c]{@{}c@{}}\textbf{Component or Block}\end{tabular}} & \begin{tabular}[r]{@{}c@{}}\textbf{Area ($mm^2$)}\end{tabular} & \textbf{\%} & \begin{tabular}[r]{@{}c@{}}\textbf{Power ($mW$)}\end{tabular}  & \textbf{\%} \\ \hline \hline
\multicolumn{2}{|l|}{{HiHGNN (4 Lanes)}}    &21.36 &100 &12001.87 &100    \\ \hline \hline
\multicolumn{6}{|c|}{\begin{tabular}[c]{@{}c@{}} \textbf{Breakdown by Functional Block} \end{tabular}} \\ \hline
\multicolumn{2}{|l|}{FP-Buf}     &1.80 &8.42 &202.90 &1.69    \\
\multicolumn{2}{|l|}{NA-Buf}   &10.70 &50.11 &1207.42 &10.06    \\
\multicolumn{2}{|l|}{SF-Buf}   &0.09 &0.41 &9.98 &0.08     \\
\multicolumn{2}{|l|}{Att-Buf}   &0.28 &1.31 &31.60 &0.26    \\
\multicolumn{2}{|l|}{Systolic Module}    &5.10 &23.88 &6758.40 &56.31   \\
\multicolumn{2}{|l|}{SIMD Module}          &3.19 &14.93 &3276.80 &27.30    \\
\multicolumn{2}{|l|}{Crossbar}          &0.09 &0.44 &440.82 &3.67    \\
\multicolumn{2}{|l|}{Others}          &0.11 &0.50 & 73.95 & 0.62    \\  \hline
\end{tabular}
}
\end{table}

\begin{figure*}[!hptb] 
    \vspace{-5pt}
    \small
    \centering
    \includegraphics[width=1.0\textwidth]{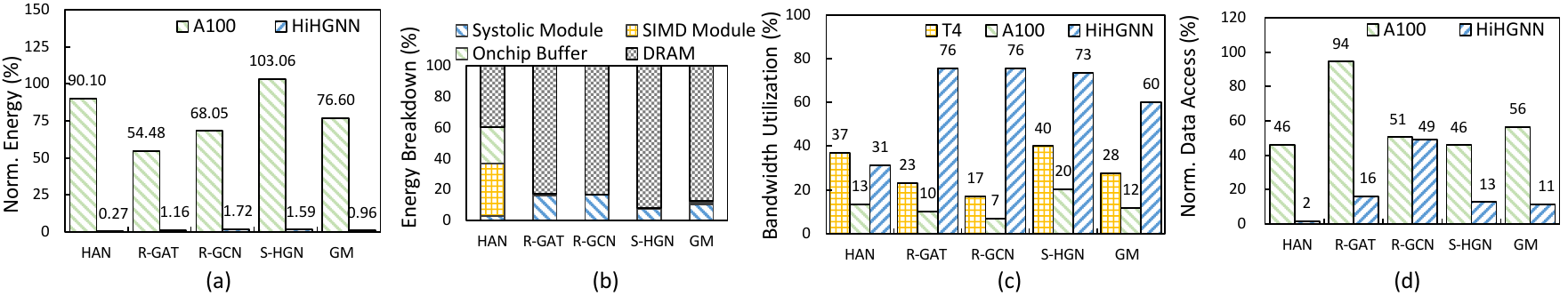}
    \vspace{-20pt}
    \caption{Details of results: (a) Normalized energy over GPU T4; (b) Energy breakdown of HiHGNN; (c) DRAM bandwidth utilization of all platforms; (d) Normalized data access to DRAM over GPU T4. We only show results on the DBLP dataset for simplicity. }
    \label{fig:overall_detailed_results}
    \vspace{-10pt}
\end{figure*}

\textbf{Energy and Its Breakdown.} Fig.~\ref{fig:overall_detailed_results} (a) shows that HiHGNN achieves the average energy reduction with 99.59\% and 99.74\% compared to GPU T4 and GPU A100, respectively. The energy consumption of all platforms includes the DRAM.
There are two factors mainly contributing to the overall reduction in energy consumption. First, the reduction of execution time leads to a decrease in energy consumption. Second, the exploitation of data reusability greatly reduces DRAM accesses, reducing energy consumption as well.

Fig.~\ref{fig:overall_detailed_results} (b) shows the breakdown of the energy consumption of different HGNNs models. 
First, the energy consumed by DRAM accesses occupies most of the energy consumption in all models. 
Second, the HAN model incurs a higher energy cost on its \textit{SIMD Module} and on-chip buffer than other models because the NA stage occupies the most execution time.
Third, both the R-GCN and R-GAT models consume a higher energy consumption in \textit{Systolic Module} compared to other models, due to the long execution time of their FP stage.
Fourth, the S-HGN model consumes more energy in the DRAM access, due to the requirement to load both raw features and the additional edge embeddings in the NA stage.

\textbf{Utilization of DRAM Bandwidth.}
Fig.~\ref{fig:overall_detailed_results} (c) shows the average utilization of the DRAM bandwidth of HiHGNN and others. 
HiHGNN reaches 2.1$\times$ of GPU T4 and 5.0$\times$ of GPU A100 in bandwidth utilization, respectively.
The bandwidth utilization for GPUs is limited by the random accesses to projected features, attention coefficients, and edge embeddings. However, those intermediate results are elaborately stored on on-chip buffers in HiHGNN, eliminating random DRAM access. It should be noted that, in the case of the HAN model, the intermediate results in the NA stage are mostly stored in NA-Buf, which significantly reduces the DRAM accesses while also reducing bandwidth utilization.

\textbf{Number of DRAM Accesses.} 
Fig.~\ref{fig:overall_detailed_results} (d) shows the total data access to the DRAM of HiHGNN and others during runtime, normalized to GPU T4. 
HiHGNN dramatically reduces DRAM accesses by 89\% and 80\% compared to GPU T4 and GPU A100 on average, respectively.
HiHGNN reduces the need for DRAM access by elaborately reusing intermediate results that are frequently accessed, such as projected features.
From the perspective of the model, the HAN model adopts the type-specific FP stage and attention-based NA stage, and is greatly benefited from the reuse of intermediate results. In addition, HAN also takes advantage of the similarity-aware execution scheduling to remove redundant data accesses at the semantic graph level, resulting in a further reduction of DRAM accesses.
On the contrary, the relation-specific FP stage of R-GCN dominates the execution, and significant DRAM accesses to raw features are inevitable.

\subsection{Effects of Proposed Optimizations} \label{sec:evaluation_optimization}

A detailed evaluation on the DBLP dataset is conducted to give more insights into the proposed optimizations. The optimizations include bound-aware stage fusion, independency-aware parallelism exploitation, and similarity-aware execution scheduling. 
We compare the following combinations to understand the effectiveness of them: 1) w/ and w/o bound-aware stage fusion; 2) the scalability of scale-up architecture as well as w/ and w/o workload-aware scheduling; 3) w/ and w/o similarity-aware execution scheduling.

\begin{figure}[!t] 
    \vspace{-5pt}
    \small
    \centering
   \includegraphics[width=0.48\textwidth]{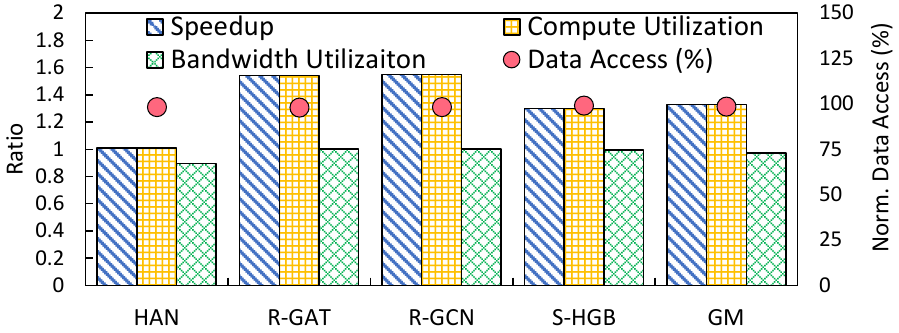}
    \vspace{-10pt}
    \caption{Performance comparison of HiHGNN w/ and w/o bound-aware stage fusion.}
    \label{fig:BPA_benefit}
    \vspace{-5pt}
\end{figure}

\textbf{Effect of Bound-aware Stage Fusion.}
Fig.~\ref{fig:BPA_benefit} depicts the benefit brought by the bound-aware stage fusion. During execution, this technique reduces 35\% execution time on average.
For the R-GCN and R-GAT models, the time-consuming FP stage greatly benefits from the bound-aware stage fusion, which results in up to 50\% improvements in compute utilization and performance. However, for HAN, the NA stage dominates the entire processing time, making the bound-aware stage fusion a limited success.

\begin{figure}[!t] 
    \vspace{-5pt}
    \small
    \centering
   \includegraphics[width=0.48\textwidth]{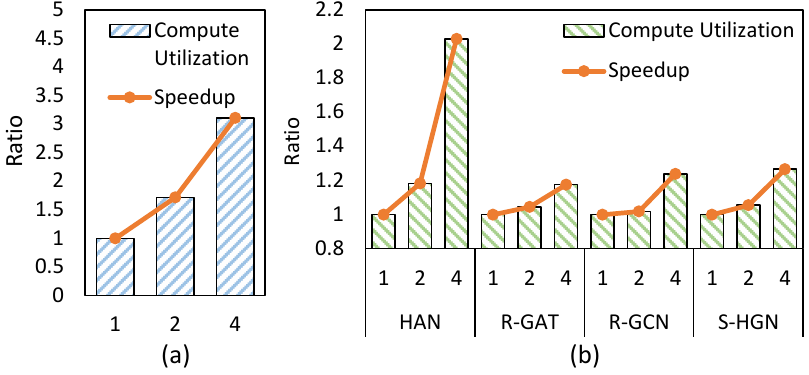}
    \vspace{-10pt}
    \caption{Performance comparison of HiHGNN w/ and w/o independency-aware parallel execution: (a) Speedup and compute utilization for a single semantic graph processing in the different number of lanes; (b) Speedup and normalized compute utilization w/ workload-aware scheduling to w/o workload-aware scheduling.}
    \label{fig:IPD_benefit}
    \vspace{-5pt}
\end{figure}

\textbf{Effect of Independency-aware Parallel Execution.}
Fig.~\ref{fig:IPD_benefit} illustrates the scalability of HiHGNN and the effect of workload-aware scheduling. 
Fig.~\ref{fig:IPD_benefit} (a) depicts an approximately linear improvement in performance and compute utilization with the increasing number of lanes.
This is because the workload-aware scheduling efficiently enables multiple lanes to work together on a single semantic graph, ensuring the scalability of the entire architecture.
Fig.~\ref{fig:IPD_benefit} (b) gives an ablation study on the impact of workload-aware scheduling. The results demonstrate that this scheduling leads to a gradual improvement in compute utilization and performance with the increasing number of lanes.
Noticed, for the HAN model, the total workload consists of three semantic graphs. Therefore, when the number of lanes scales up to four, this scheduling allows for an additional idle lane to be utilized for processing, resulting in a significant improvement.

\begin{figure}[!t] 
    \vspace{-5pt}
    \small
    \centering
    \includegraphics[width=0.48\textwidth]{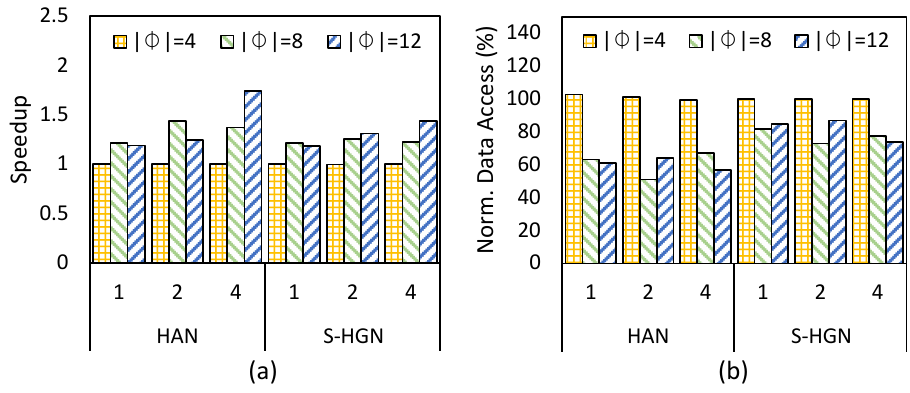}
    \vspace{-10pt}
    \caption{Performance of HiHGNN w/ and w/o similarity-aware execution scheduling: (a) Speedup and (b) normalized DRAM data access across different ratios and semantic graphs.}
    \label{fig:SES_benefit}
    \vspace{-5pt}
\end{figure}

\textbf{Effect of Similarity-aware Execution Scheduling.}
Fig.~\ref{fig:SES_benefit} (a) and (b) show the speedup and DRAM data access in the four-lane architecture of HiHGNN with the similarity-aware execution scheduling, respectively, compared to the random scheduling case. Note that the horizontal axis represents the ratio between the size of total projected vertex features and the size of the FP-Buf.

In detail, when the number of semantic graphs is four, the scheduling has limited impact on performance and DRAM data access because of the four-lane architecture. However, as the number of semantic graphs increases to eight and twelve, the scheduling demonstrates better performance improvement and more reduction of DRAM access. This improvement is attributed to the effective reuse of projected features from previous semantic graph by the subsequent one. Furthermore, the prepossessing overhead of this scheduling on CPU is negligible. For instance, on the DBLP dataset, it accounts for less than 0.1\% of the end-to-end execution time.


\subsection{Results on FPGA Implementation} 

\begin{figure}[!t] 
    \small
    \centering
    \includegraphics[width=0.48\textwidth]{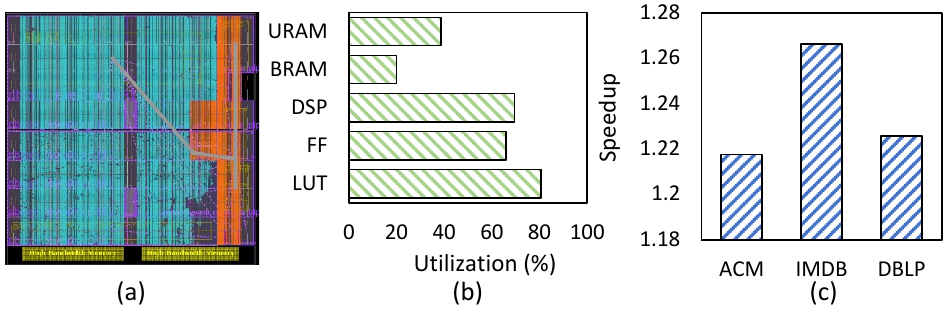}
    \vspace{-10pt}
    \caption{Results on FPGA implementation for a small single lane design: (a) Layout of HiHGNN; (b) Various resources utilization ratios; (c) Speedup of HiHGNN w/ and w/o similarity-aware execution scheduling.}
    \label{fig:FPGA}
    \vspace{-5pt}
\end{figure}

To further validate the feasibility of our design, we implement and evaluate a small single-lane design of HiHGNN on the Xilinx FPGA Alveo U50 accelerator card using Xilinx Vitis Toolchains. Note that due to the limited arithmetic and bandwidth resources on FPGA board, our FPGA prototype implementation is only used to validate the implementability of the architecture and the effectiveness of the strategy.

Fig.~\ref{fig:FPGA} (a) and (b) respectively present the implementation layout and utilization ratios of various types of resources of FPGA. 
Fig.~\ref{fig:FPGA} (c) shows that HiHGNN achieves an average speedup of 1.23$\times$ among various datasets with the adoption of the similarity-aware execution scheduling. This result is consistent with the result in the previous subsection.

\section{Related Work}
\label{sec:related_work}

Due to the outstanding learning capacity of GNNs on graph data, GNN accelerators draw great interest from the architecture community~\cite{HyGCN, awbgcn,igcn, GRIP, FlowGNN, EnGN, ReGNN, GROW, SGCN, REFLIP-HUAKE, ReGNN-HUAKE, GCNAX, Hyperscale, FPGAN, NTGAT, FTW-GAT, GCN_Bidirectional_Fusion, MultiGCN}.

One part of them~\cite{HyGCN,GCoD,GRIP} employs stage fusion to boost the overall performance. 
For example, \textit{HyGCN}~\cite{HyGCN} proposes a hybrid architecture to leverage inter-stage fusion to improve overall performance. 
\textit{GCoD}~\cite{GCoD} proposes a GCN algorithm and accelerator co-design framework dedicated two-pronged accelerator with a separate engine to process each of the denser and sparser workloads. Both of the two engines achieve the inter-stage fusion.
\textit{GRIP}~\cite{GRIP} decomposes GNN inference into three stages and implements each phase with specialized on-chip memory and execution units, realizing the fusion of these three stages.

The aforementioned accelerators can be categorized as having a design based on parallel pipeline (PP), according to the definition in the literature~\cite{GNN_dataflow_taxonomy}. In a PP design, the parallel units (i.e., groups of PEs) within the accelerator execute both stages simultaneously. However, the workflow of HGNN involves more intricate execution processes and additional execution stages, resulting in different fusion methodology.
Another part of them~\cite{FlowGNN,EnGN,ReGNN} improves performance by exploiting multi-dimensional parallelism. For example, \textit{FlowGNN}~\cite{FlowGNN} uses a scalable dataflow architecture to generally support a wide range of GNN models and exploits multiple-level parallelism in terms of the processing of vertices and edges. 
Unfortunately, previous efforts are ill-suited to exploit the high-degree parallelism and data reusability which are inherent in HGNNs. 
In addition, most HGNNs leverage attention mechanisms to improve model accuracy. However, most GNN accelerators focus on graph convolutional network (GCN)~\cite{GCN} and GCN variants, failing to perform attention mechanisms. 
A few other efforts~\cite{NTGAT,FTW-GAT,FPGAN} using parameter quantization and model pruning that sacrifice accuracy to accelerate graph attention networks. Their optimizations are orthogonal to ours.

The first DIMM-based near-memory processing HGNNs accelerator, MetaNMP, is recently proposed, greatly reducing memory footprint and eliminate redundant computations~\cite{MetaNMP}. 
The former is achieved by employing a cartesian-like product paradigm to generate all metapath instances on the fly. While for the latter, it aggregates vertex features along the dispersed metapath instances from the starting vertex, effectively utilizing shareable aggregation computations. 
MetaNMP mainly focuses on the HGNN model which uses the intra-metapath aggregation~\cite{MAGNN} to aggregate vertex features along the metapath in feature projection stage. However, the intra-metapath aggregation is not widely used in the algorithm community after proposed by MAGNN~\cite{MAGNN} in 2020.
Unlike MetaNMP, our work aims to exploit the inter-semantic-graph parallelism and data reusability in HGNNs, which is universal in HGNNs models.
In addition, based on the DIMM-based near-memory processing, MetaNMP cannot be adopted as a near term solution since this memory technology is still immature. In contrast, our work presents a practical solution by leveraging the off-the-shelf HBM.


A number of previous efforts characterize the execution of HGNNs~\cite{understand_HGNN} and GNNs~\cite{understand_GCN,understand_distributed_gnn,GNNMark,understand_and_bridge,understand_gnn_computational_graph,analysis_of_bottlenecks,characterize_gnn_accelerators,in_depth_concurrency_analysis} in detail for the software and hardware optimizations. Our work~\cite{understand_HGNN} quantitatively characterizes the inference phase of HGNNs on GPUs, in which we reveal the performance bottleneck, execution pattern, and execution semantic of HGNNs. Our other work~\cite{understand_GCN,understand_distributed_gnn} characterizes the hybrid execution pattern of GCNs on GPUs and distributed GNN training on a multi-node GPU system. Additionally, GNNMark~\cite{GNNMark} introduces a unified evaluation framework that supports various GNN models and datasets, enabling performance bottleneck analysis and evaluation of system-level metrics such as scalability.


DGL~\cite{DGL} and PyG~\cite{PyG} are two popular GPU-based Python libraries used for the HGNN models implementation. For example, DGL provides many high-performance GPU kernels to perform various operations, such as the SpMMCsr kernel for the NA stage. However, GPU inherently suffers from low efficiency in the performing of irregular memory accesses, e.g., encountering massive unavailing replacements of features during the NA stage.

\section{conclusion}
In this paper, a set of well-known HGNN models are quantitatively characterized on GPU to disclose the execution patterns, performance bottlenecks, and acceleration opportunities of HGNNs.
Then, a high-performance HGNN accelerator, HiHGNN, is designed to exploit the high-degree parallelism and data reusability in HGNNs. HiHGNN achieves higher performance and energy efficiency compared with the high-end GPU A100. We believe that our findings and design will attract more scholars to participate in designing the architecture for increasingly important HGNNs.



\ifCLASSOPTIONcompsoc
    \section*{Acknowledgments}
\else
    \section*{Acknowledgment}
\fi

This work was supported by National Key Research and Development Program (Grant No. 2022YFB4501400), the National Natural Science Foundation of China (Grant No. 62202451), CAS Project for Young Scientists in Basic Research (Grant No. YSBR-029), CAS Project for Youth Innovation Promotion Association, Beijing Natural Science Foundation (Grant No. L234078) and Beijing Nova Program (Grant No. 20220484054, No.20230484420).


\ifCLASSOPTIONcaptionsoff
  \newpage
\fi



%



\bibliographystyle{IEEEtranS}
\bibliography{refs}

\begin{thebibliography}{10}
\providecommand{\url}[1]{#1}
\csname url@samestyle\endcsname
\providecommand{\newblock}{\relax}
\providecommand{\bibinfo}[2]{#2}
\providecommand{\BIBentrySTDinterwordspacing}{\spaceskip=0pt\relax}
\providecommand{\BIBentryALTinterwordstretchfactor}{4}
\providecommand{\BIBentryALTinterwordspacing}{\spaceskip=\fontdimen2\font plus
\BIBentryALTinterwordstretchfactor\fontdimen3\font minus \fontdimen4\font\relax}
\providecommand{\BIBforeignlanguage}[2]{{%
\expandafter\ifx\csname l@#1\endcsname\relax
\typeout{** WARNING: IEEEtranS.bst: No hyphenation pattern has been}%
\typeout{** loaded for the language `#1'. Using the pattern for}%
\typeout{** the default language instead.}%
\else
\language=\csname l@#1\endcsname
\fi
#2}}
\providecommand{\BIBdecl}{\relax}
\BIBdecl

\bibitem{CACTI}
\BIBentryALTinterwordspacing
Cacti. [Online]. Available: \url{http://www.hpl.hp.com/research/cacti/}
\BIBentrySTDinterwordspacing

\bibitem{Tesseract}
J.~Ahn, S.~Hong, S.~Yoo, O.~Mutlu, and K.~Choi, ``A scalable processing-in-memory accelerator for parallel graph processing,'' in \emph{Proceedings of the 42nd Annual International Symposium on Computer Architecture}, 2015, pp. 105--117.

\bibitem{A2N}
T.~Bansal, D.-C. Juan, S.~Ravi, and A.~McCallum, ``A2n: Attending to neighbors for knowledge graph inference,'' in \emph{Proceedings of the 57th annual meeting of the association for computational linguistics}, 2019, pp. 4387--4392.

\bibitem{GNNMark}
T.~Baruah, K.~Shivdikar, S.~Dong \emph{et~al.}, ``Gnnmark: A benchmark suite to characterize graph neural network training on gpus,'' in \emph{2021 IEEE International Symposium on Performance Analysis of Systems and Software (ISPASS)}, 2021, pp. 13--23.

\bibitem{in_depth_concurrency_analysis}
M.~Besta and T.~Hoefler, ``Parallel and distributed graph neural networks: An in-depth concurrency analysis,'' \emph{arXiv preprint arXiv:2205.09702}, 2022.

\bibitem{workloadbalance}
\BIBentryALTinterwordspacing
R.~D. Blumofe and C.~E. Leiserson, ``Scheduling multithreaded computations by work stealing,'' \emph{J. ACM}, vol.~46, no.~5, p. 720–748, sep 1999. [Online]. Available: \url{https://doi.org/10.1145/324133.324234}
\BIBentrySTDinterwordspacing

\bibitem{ReGNN}
C.~Chen, K.~Li, Y.~Li, and X.~Zou, ``Regnn: A redundancy-eliminated graph neural networks accelerator,'' in \emph{2022 IEEE International Symposium on High-Performance Computer Architecture (HPCA)}, 2022, pp. 429--443.

\bibitem{MetaNMP}
\BIBentryALTinterwordspacing
D.~Chen, H.~He, H.~Jin \emph{et~al.}, ``Metanmp: Leveraging cartesian-like product to accelerate hgnns with near-memory processing,'' in \emph{Proceedings of the 50th Annual International Symposium on Computer Architecture}, ser. ISCA '23.\hskip 1em plus 0.5em minus 0.4em\relax New York, NY, USA: Association for Computing Machinery, 2023. [Online]. Available: \url{https://doi.org/10.1145/3579371.3589091}
\BIBentrySTDinterwordspacing

\bibitem{chen2017task}
T.~Chen and Y.~Sun, ``Task-guided and path-augmented heterogeneous network embedding for author identification,'' in \emph{Proceedings of the tenth ACM international conference on web search and data mining}, 2017, pp. 295--304.

\bibitem{hgnn_survey_tangjie}
Y.~Dong, Z.~Hu, K.~Wang, Y.~Sun, and J.~Tang, ``Heterogeneous network representation learning.'' in \emph{IJCAI}, vol.~20, 2020, pp. 4861--4867.

\bibitem{fan2019metapath}
S.~Fan, J.~Zhu, X.~Han \emph{et~al.}, ``Metapath-guided heterogeneous graph neural network for intent recommendation,'' in \emph{Proceedings of the 25th ACM SIGKDD international conference on knowledge discovery \& data mining}, 2019, pp. 2478--2486.

\bibitem{PyG}
M.~Fey and J.~E. Lenssen, ``Fast graph representation learning with {PyTorch Geometric},'' in \emph{ICLR Workshop on Representation Learning on Graphs and Manifolds}, 2019.

\bibitem{MAGNN}
X.~Fu, J.~Zhang, Z.~Meng, and I.~King, ``Magnn: Metapath aggregated graph neural network for heterogeneous graph embedding,'' in \emph{Proceedings of The Web Conference 2020}, 2020, pp. 2331--2341.

\bibitem{GNN_dataflow_taxonomy}
R.~Garg, E.~Qin, F.~Muñoz-Matrínez, R.~Guirado, A.~Jain, S.~Abadal, J.~L. Abellán, M.~E. Acacio, E.~Alarcón, S.~Rajamanickam, and T.~Krishna, ``Understanding the design-space of sparse/dense multiphase gnn dataflows on spatial accelerators,'' in \emph{2022 IEEE International Parallel and Distributed Processing Symposium (IPDPS)}, 2022, pp. 571--582.

\bibitem{awbgcn}
T.~Geng, A.~Li, R.~Shi \emph{et~al.}, ``Awb-gcn: A graph convolutional network accelerator with runtime workload rebalancing,'' in \emph{2020 53rd Annual IEEE/ACM International Symposium on Microarchitecture (MICRO)}, 2020, pp. 922--936.

\bibitem{igcn}
\BIBentryALTinterwordspacing
T.~Geng, C.~Wu, Y.~Zhang \emph{et~al.}, ``I-gcn: A graph convolutional network accelerator with runtime locality enhancement through islandization,'' in \emph{MICRO-54: 54th Annual IEEE/ACM International Symposium on Microarchitecture}, ser. MICRO '21.\hskip 1em plus 0.5em minus 0.4em\relax New York, NY, USA: Association for Computing Machinery, 2021, p. 1051–1063. [Online]. Available: \url{https://doi.org/10.1145/3466752.3480113}
\BIBentrySTDinterwordspacing

\bibitem{HGL}
Y.~Gui, Y.~Wu, H.~Yang \emph{et~al.}, ``Hgl: Accelerating heterogeneous gnn training with holistic representation and optimization,'' in \emph{SC22: International Conference for High Performance Computing, Networking, Storage and Analysis}, 2022, pp. 1--15.

\bibitem{characterize_gnn_accelerators}
R.~Guirado, A.~Jain, S.~Abadal, and E.~Alarcón, ``Characterizing the communication requirements of gnn accelerators: A model-based approach,'' in \emph{2021 IEEE International Symposium on Circuits and Systems (ISCAS)}, 2021, pp. 1--5.

\bibitem{Graphicionado}
T.~J. Ham, L.~Wu, N.~Sundaram, N.~Satish, and M.~Martonosi, ``Graphicionado: {{A}} high-performance and energy-efficient accelerator for graph analytics,'' in \emph{2016 49th {{Annual IEEE}}/{{ACM International Symposium}} on {{Microarchitecture}} ({{MICRO}})}, Oct. 2016, pp. 1--13.

\bibitem{FTW-GAT}
Z.~He, T.~Tian, Q.~Wu, and X.~Jin, ``Ftw-gat: An fpga-based accelerator for graph attention networks with ternary weights,'' \emph{IEEE Transactions on Circuits and Systems II: Express Briefs}, pp. 1--1, 2023.

\bibitem{NTGAT}
\BIBentryALTinterwordspacing
W.~Hou, K.~Zhong, S.~Zeng, G.~Dai, H.~Yang, and Y.~Wang, ``Ntgat: A graph attention network accelerator with runtime node tailoring,'' in \emph{Proceedings of the 28th Asia and South Pacific Design Automation Conference}, ser. ASPDAC '23.\hskip 1em plus 0.5em minus 0.4em\relax New York, NY, USA: Association for Computing Machinery, 2023, p. 645–650. [Online]. Available: \url{https://doi.org/10.1145/3566097.3567869}
\BIBentrySTDinterwordspacing

\bibitem{understand_and_bridge}
\BIBentryALTinterwordspacing
K.~Huang, J.~Zhai, Z.~Zheng, Y.~Yi, and X.~Shen, ``Understanding and bridging the gaps in current gnn performance optimizations,'' in \emph{Proceedings of the 26th ACM SIGPLAN Symposium on Principles and Practice of Parallel Programming}, ser. PPoPP '21.\hskip 1em plus 0.5em minus 0.4em\relax New York, NY, USA: Association for Computing Machinery, 2021, p. 119–132. [Online]. Available: \url{https://doi.org/10.1145/3437801.3441585}
\BIBentrySTDinterwordspacing

\bibitem{REFLIP-HUAKE}
\BIBentryALTinterwordspacing
Y.~Huang, L.~Zheng, P.~Yao \emph{et~al.}, ``Accelerating graph convolutional networks using crossbar-based processing-in-memory architectures,'' in \emph{{IEEE} International Symposium on High-Performance Computer Architecture, {HPCA} 2022, Seoul, South Korea, April 2-6, 2022}.\hskip 1em plus 0.5em minus 0.4em\relax {IEEE}, 2022, pp. 1029--1042. [Online]. Available: \url{https://doi.org/10.1109/HPCA53966.2022.00079}
\BIBentrySTDinterwordspacing

\bibitem{GROW}
\BIBentryALTinterwordspacing
R.~Hwang, M.~Kang, J.~Lee, D.~Kam, Y.~Lee, and M.~Rhu, ``{GROW:} {A} row-stationary sparse-dense {GEMM} accelerator for memory-efficient graph convolutional neural networks,'' in \emph{{IEEE} International Symposium on High-Performance Computer Architecture, {HPCA} 2023, Montreal, QC, Canada, February 25 - March 1, 2023}.\hskip 1em plus 0.5em minus 0.4em\relax {IEEE}, 2023, pp. 42--55. [Online]. Available: \url{https://doi.org/10.1109/HPCA56546.2023.10070983}
\BIBentrySTDinterwordspacing

\bibitem{TPU}
N.~P. Jouppi, C.~Young, N.~Patil \emph{et~al.}, ``In-datacenter performance analysis of a tensor processing unit,'' in \emph{Proceedings of the 44th Annual International Symposium on Computer Architecture}, ser. {ISCA} '17.\hskip 1em plus 0.5em minus 0.4em\relax {ACM}, pp. 1--12.

\bibitem{ramulator}
Y.~Kim, W.~Yang, and O.~Mutlu, ``Ramulator: A fast and extensible dram simulator,'' \emph{IEEE Computer architecture letters}, vol.~15, no.~1, pp. 45--49, 2015.

\bibitem{GRIP}
K.~Kiningham, P.~Levis, and C.~Ré, ``Grip: A graph neural network accelerator architecture,'' \emph{IEEE Transactions on Computers}, vol.~72, no.~4, pp. 914--925, 2023.

\bibitem{GCN}
\BIBentryALTinterwordspacing
T.~N. Kipf and M.~Welling, ``Semi-supervised classification with graph convolutional networks,'' in \emph{International Conference on Learning Representations, {ICLR} 2017}, 2017. [Online]. Available: \url{https://openreview.net/forum?id=SJU4ayYgl}
\BIBentrySTDinterwordspacing

\bibitem{li2022disentangled}
A.~Li, Z.~Cheng, F.~Liu, Z.~Gao, W.~Guan, and Y.~Peng, ``Disentangled graph neural networks for session-based recommendation,'' \emph{IEEE Transactions on Knowledge and Data Engineering}, 2022.

\bibitem{GCN_Bidirectional_Fusion}
H.~Li, M.~Yan, X.~Yang \emph{et~al.}, ``Hardware acceleration for gcns via bidirectional fusion,'' \emph{IEEE Computer Architecture Letters}, vol.~20, no.~1, pp. 66--4, 2021.

\bibitem{GCNAX}
\BIBentryALTinterwordspacing
J.~Li, A.~Louri, A.~Karanth, and R.~C. Bunescu, ``{GCNAX:} {A} flexible and energy-efficient accelerator for graph convolutional neural networks,'' in \emph{{IEEE} International Symposium on High-Performance Computer Architecture, {HPCA} 2021, Seoul, South Korea, February 27 - March 3, 2021}.\hskip 1em plus 0.5em minus 0.4em\relax {IEEE}, 2021, pp. 775--788. [Online]. Available: \url{https://doi.org/10.1109/HPCA51647.2021.00070}
\BIBentrySTDinterwordspacing

\bibitem{Hyperscale}
\BIBentryALTinterwordspacing
S.~Li, D.~Niu, Y.~Wang \emph{et~al.}, ``Hyperscale fpga-as-a-service architecture for large-scale distributed graph neural network,'' in \emph{Proceedings of the 49th Annual International Symposium on Computer Architecture}, ser. ISCA '22.\hskip 1em plus 0.5em minus 0.4em\relax New York, NY, USA: Association for Computing Machinery, 2022, p. 946–961. [Online]. Available: \url{https://doi.org/10.1145/3470496.3527439}
\BIBentrySTDinterwordspacing

\bibitem{EnGN}
S.~Liang, Y.~Wang, C.~Liu \emph{et~al.}, ``Engn: A high-throughput and energy-efficient accelerator for large graph neural networks,'' \emph{IEEE Trans. Comput.}, vol.~70, no.~9, p. 1511–1525, sep 2021.

\bibitem{understand_distributed_gnn}
H.~Lin, M.~Yan, X.~Yang \emph{et~al.}, ``Characterizing and understanding distributed gnn training on gpus,'' \emph{IEEE Computer Architecture Letters}, vol.~21, no.~1, pp. 21--24, 2022.

\bibitem{Comprehensive_Survey_GNN_Distributed_Training}
H.~Lin, M.~Yan, X.~Ye \emph{et~al.}, ``A comprehensive survey on distributed training of graph neural networks,'' \emph{arXiv preprint arXiv:2211.05368}, 2022.

\bibitem{gpu_drawback}
E.~Lindholm, J.~Nickolls, S.~Oberman, and J.~Montrym, ``Nvidia tesla: A unified graphics and computing architecture,'' \emph{IEEE micro}, vol.~28, no.~2, pp. 39--55, 2008.

\bibitem{ReGNN-HUAKE}
\BIBentryALTinterwordspacing
C.~Liu, H.~Liu, H.~Jin \emph{et~al.}, ``Regnn: a reram-based heterogeneous architecture for general graph neural networks,'' in \emph{{DAC} '22: 59th {ACM/IEEE} Design Automation Conference, San Francisco, California, USA, July 10 - 14, 2022}, R.~Oshana, Ed.\hskip 1em plus 0.5em minus 0.4em\relax {ACM}, 2022, pp. 469--474. [Online]. Available: \url{https://doi.org/10.1145/3489517.3530479}
\BIBentrySTDinterwordspacing

\bibitem{liu2018heterogeneous}
Z.~Liu, C.~Chen, X.~Yang, J.~Zhou, X.~Li, and L.~Song, ``Heterogeneous graph neural networks for malicious account detection,'' in \emph{Proceedings of the 27th ACM international conference on information and knowledge management}, 2018, pp. 2077--2085.

\bibitem{luo2021imas}
F.~Luo, Y.~Zhang, and X.~Wang, ``Imas++ an intelligent medical analysis system enhanced with deep graph neural networks,'' in \emph{Proceedings of the 30th ACM International Conference on Information \& Knowledge Management}, 2021, pp. 4754--4758.

\bibitem{Simple-HGN}
Q.~Lv, M.~Ding, Q.~Liu \emph{et~al.}, ``Are we really making much progress? revisiting, benchmarking and refining heterogeneous graph neural networks,'' in \emph{Proceedings of the 27th ACM SIGKDD Conference on Knowledge Discovery \& Data Mining}, 2021, pp. 1150--1160.

\bibitem{mao2020item}
K.~Mao, X.~Xiao, J.~Zhu, B.~Lu, R.~Tang, and X.~He, ``Item tagging for information retrieval: a tripartite graph neural network based approach,'' in \emph{Proceedings of the 43rd International ACM SIGIR Conference on Research and Development in Information Retrieval}, 2020, pp. 2327--2336.

\bibitem{niu2020dual}
X.~Niu, B.~Li, C.~Li \emph{et~al.}, ``A dual heterogeneous graph attention network to improve long-tail performance for shop search in e-commerce,'' in \emph{Proceedings of the 26th ACM SIGKDD International Conference on Knowledge Discovery \& Data Mining}, 2020, pp. 3405--3415.

\bibitem{oh2018knowledge}
B.~Oh, S.~Seo, and K.-H. Lee, ``Knowledge graph completion by context-aware convolutional learning with multi-hop neighborhoods,'' in \emph{Proceedings of the 27th ACM International Conference on Information and Knowledge Management}, 2018, pp. 257--266.

\bibitem{ozdal_energy_2016}
M.~M. Ozdal, S.~Yesil, T.~Kim \emph{et~al.}, ``Energy efficient architecture for graph analytics accelerators,'' in \emph{2016 {ACM}/{IEEE} 43rd Annual International Symposium on Computer Architecture ({ISCA})}, pp. 166--177.

\bibitem{7pj}
M.~O’Connor, ``Highlights of the high-bandwidth memory (hbm) standard,'' in \emph{Memory forum workshop}, vol.~3, 2014.

\bibitem{FlowGNN}
R.~Sarkar, S.~Abi-Karam, Y.~He, L.~Sathidevi, and C.~Hao, ``Flowgnn: A dataflow architecture for real-time workload-agnostic graph neural network inference,'' in \emph{2023 IEEE International Symposium on High-Performance Computer Architecture (HPCA)}.\hskip 1em plus 0.5em minus 0.4em\relax Los Alamitos, CA, USA: IEEE Computer Society, mar 2023, pp. 1099--1112.

\bibitem{R-GCN}
M.~Schlichtkrull, T.~N. Kipf, P.~Bloem, R.~v.~d. Berg, I.~Titov, and M.~Welling, ``Modeling relational data with graph convolutional networks,'' in \emph{European semantic web conference}.\hskip 1em plus 0.5em minus 0.4em\relax Springer, 2018, pp. 593--607.

\bibitem{HG_survey}
C.~Shi, Y.~Li, J.~Zhang, Y.~Sun, and S.~Y. Philip, ``A survey of heterogeneous information network analysis,'' \emph{IEEE Transactions on Knowledge and Data Engineering}, vol.~29, no.~1, pp. 17--37, 2016.

\bibitem{socher2013reasoning}
R.~Socher, D.~Chen, C.~D. Manning, and A.~Ng, ``Reasoning with neural tensor networks for knowledge base completion,'' \emph{Advances in neural information processing systems}, vol.~26, 2013.

\bibitem{GraphR}
L.~Song, Y.~Zhuo, X.~Qian, H.~Li, and Y.~Chen, ``Graphr: Accelerating graph processing using reram,'' in \emph{2018 IEEE International Symposium on High Performance Computer Architecture (HPCA)}.\hskip 1em plus 0.5em minus 0.4em\relax IEEE, 2018, pp. 531--543.

\bibitem{MultiGCN}
G.~Sun, M.~Yan, D.~Wang \emph{et~al.}, ``Multi-node acceleration for large-scale gcns,'' \emph{IEEE Transactions on Computers}, vol.~71, no.~12, pp. 3140--3152, 2022.

\bibitem{sun2012mining}
Y.~Sun and J.~Han, ``Mining heterogeneous information networks: principles and methodologies,'' \emph{Synthesis Lectures on Data Mining and Knowledge Discovery}, vol.~3, no.~2, pp. 1--159, 2012.

\bibitem{sun2011pathsim}
Y.~Sun, J.~Han, X.~Yan, P.~S. Yu, and T.~Wu, ``Pathsim: Meta path-based top-k similarity search in heterogeneous information networks,'' \emph{Proceedings of the VLDB Endowment}, vol.~4, no.~11, pp. 992--1003, 2011.

\bibitem{tajeuna2018modeling}
E.~G. Tajeuna, M.~Bouguessa, and S.~Wang, ``Modeling and predicting community structure changes in time-evolving social networks,'' \emph{IEEE Transactions on Knowledge and Data Engineering}, vol.~31, no.~6, pp. 1166--1180, 2018.

\bibitem{CompGCN}
\BIBentryALTinterwordspacing
S.~Vashishth, S.~Sanyal, V.~Nitin, and P.~Talukdar, ``Composition-based multi-relational graph convolutional networks,'' in \emph{International Conference on Learning Representations}, 2020. [Online]. Available: \url{https://openreview.net/forum?id=BylA_C4tPr}
\BIBentrySTDinterwordspacing

\bibitem{GAT}
\BIBentryALTinterwordspacing
P.~Veli{\v{c}}kovi{\'{c}}, G.~Cucurull, A.~Casanova, A.~Romero, P.~Li{\`{o}}, and Y.~Bengio, ``{Graph Attention Networks},'' \emph{International Conference on Learning Representations, {ICLR} 2018}, 2018. [Online]. Available: \url{https://openreview.net/forum?id=rJXMpikCZ}
\BIBentrySTDinterwordspacing

\bibitem{technology_scale}
O.~{Villa}, D.~R. {Johnson}, M.~{Oconnor} \emph{et~al.}, ``Scaling the power wall: A path to exascale,'' in \emph{SC '14: Proceedings of the International Conference for High Performance Computing, Networking, Storage and Analysis}, Nov 2014, pp. 830--841.

\bibitem{R-GAT}
K.~Wang, W.~Shen, Y.~Yang, X.~Quan, and R.~Wang, ``Relational graph attention network for aspect-based sentiment analysis,'' in \emph{Proceedings of the 58th Annual Meeting of the Association for Computational Linguistics}, 2020, pp. 3229--3238.

\bibitem{DGL}
M.~Y. Wang, ``Deep graph library: Towards efficient and scalable deep learning on graphs,'' in \emph{ICLR workshop on representation learning on graphs and manifolds}, 2019.

\bibitem{M2GNN}
S.~Wang, X.~Wei, C.~N. Nogueira~dos Santos \emph{et~al.}, ``Mixed-curvature multi-relational graph neural network for knowledge graph completion,'' in \emph{Proceedings of the Web Conference 2021}, 2021, pp. 1761--1771.

\bibitem{hgnn_survey_shichuan}
X.~Wang, D.~Bo, C.~Shi, S.~Fan, Y.~Ye, and P.~S. Yu, ``A survey on heterogeneous graph embedding: methods, techniques, applications and sources,'' \emph{arXiv preprint arXiv:2011.14867}, 2020.

\bibitem{HAN}
X.~Wang, H.~Ji, C.~Shi \emph{et~al.}, ``Heterogeneous graph attention network,'' in \emph{The world wide web conference}, 2019, pp. 2022--2032.

\bibitem{analysis_of_bottlenecks}
\BIBentryALTinterwordspacing
Z.~Wang, Y.~Wang, C.~Yuan, R.~Gu, and Y.~Huang, ``Empirical analysis of performance bottlenecks in graph neural network training and inference with gpus,'' \emph{Neurocomputing}, vol. 446, pp. 165--191, 2021. [Online]. Available: \url{https://www.sciencedirect.com/science/article/pii/S0925231221003659}
\BIBentrySTDinterwordspacing

\bibitem{comprehensive_gnn_survey}
Z.~Wu, S.~Pan, F.~Chen, G.~Long, C.~Zhang, and S.~Y. Philip, ``A comprehensive survey on graph neural networks,'' \emph{IEEE transactions on neural networks and learning systems}, vol.~32, no.~1, pp. 4--24, 2020.

\bibitem{understand_GCN}
M.~Yan, Z.~Chen, L.~Deng \emph{et~al.}, ``Characterizing and understanding gcns on gpu,'' \emph{IEEE Computer Architecture Letters}, vol.~19, no.~1, pp. 22--25, 2020.

\bibitem{HyGCN}
M.~Yan, L.~Deng, X.~Hu \emph{et~al.}, ``Hygcn: A gcn accelerator with hybrid architecture,'' in \emph{2020 IEEE International Symposium on High Performance Computer Architecture (HPCA)}.\hskip 1em plus 0.5em minus 0.4em\relax IEEE, 2020, pp. 15--29.

\bibitem{GraphDynS}
\BIBentryALTinterwordspacing
M.~Yan, X.~Hu, S.~Li \emph{et~al.}, ``Alleviating irregularity in graph analytics acceleration: A hardware/software co-design approach,'' in \emph{Proceedings of the 52nd Annual IEEE/ACM International Symposium on Microarchitecture}, ser. MICRO '52.\hskip 1em plus 0.5em minus 0.4em\relax New York, NY, USA: Association for Computing Machinery, 2019, p. 615–628. [Online]. Available: \url{https://doi.org/10.1145/3352460.3358318}
\BIBentrySTDinterwordspacing

\bibitem{understand_HGNN}
M.~Yan, M.~Zou, X.~Yang \emph{et~al.}, ``Characterizing and understanding hgnns on gpus,'' \emph{IEEE Computer Architecture Letters}, vol.~21, no.~2, pp. 69--72, 2022.

\bibitem{FPGAN}
W.~Yan, W.~Tong, and X.~Zhi, ``Fpgan: An fpga accelerator for graph attention networks with software and hardware co-optimization,'' \emph{IEEE Access}, vol.~8, pp. 171\,608--171\,620, 2020.

\bibitem{hgnn_survey_hanjiawei}
C.~Yang, Y.~Xiao, Y.~Zhang, Y.~Sun, and J.~Han, ``Heterogeneous network representation learning: A unified framework with survey and benchmark,'' \emph{IEEE Transactions on Knowledge and Data Engineering}, 2020.

\bibitem{SeHGNN}
X.~Yang, M.~Yan, S.~Pan, X.~Ye, and D.~Fan, ``Simple and efficient heterogeneous graph neural network,'' in \emph{Proceedings of the AAAI Conference on Artificial Intelligence}, vol.~37, 2023.

\bibitem{yasunaga2019scisummnet}
M.~Yasunaga, J.~Kasai, R.~Zhang \emph{et~al.}, ``Scisummnet: A large annotated corpus and content-impact models for scientific paper summarization with citation networks,'' in \emph{Proceedings of the AAAI Conference on Artificial Intelligence}, vol.~33, no.~01, 2019, pp. 7386--7393.

\bibitem{SGCN}
\BIBentryALTinterwordspacing
M.~Yoo, J.~Song, J.~Lee, N.~Kim, Y.~Kim, and J.~Lee, ``Sgcn: Exploiting compressed-sparse features in deep graph convolutional network accelerators,'' in \emph{2023 IEEE International Symposium on High-Performance Computer Architecture (HPCA)}.\hskip 1em plus 0.5em minus 0.4em\relax Los Alamitos, CA, USA: IEEE Computer Society, mar 2023, pp. 1--14. [Online]. Available: \url{https://doi.ieeecomputersociety.org/10.1109/HPCA56546.2023.10071102}
\BIBentrySTDinterwordspacing

\bibitem{GCoD}
H.~You, T.~Geng, Y.~Zhang, A.~Li, and Y.~Lin, ``Gcod: Graph convolutional network acceleration via dedicated algorithm and accelerator co-design,'' in \emph{2022 IEEE International Symposium on High-Performance Computer Architecture (HPCA)}, 2022, pp. 460--474.

\bibitem{understand_gnn_computational_graph}
H.~Zhang, Z.~Yu, G.~Dai \emph{et~al.}, ``Understanding gnn computational graph: A coordinated computation, io, and memory perspective,'' 2021.

\bibitem{zhang2019iteratively}
W.~Zhang, B.~Paudel, L.~Wang \emph{et~al.}, ``Iteratively learning embeddings and rules for knowledge graph reasoning,'' in \emph{The World Wide Web Conference}, 2019, pp. 2366--2377.

\bibitem{DNNBuilder}
X.~Zhang, J.~Wang, C.~Zhu \emph{et~al.}, ``Dnnbuilder: an automated tool for building high-performance dnn hardware accelerators for fpgas,'' in \emph{2018 IEEE/ACM International Conference on Computer-Aided Design (ICCAD)}, 2018, pp. 1--8.

\bibitem{hgnn_survey_shiruipan}
X.~Zheng, Y.~Liu, S.~Pan, M.~Zhang, D.~Jin, and P.~S. Yu, ``Graph neural networks for graphs with heterophily: A survey,'' \emph{arXiv preprint arXiv:2202.07082}, 2022.

\bibitem{zheng2020clustering}
Y.~Zheng, R.~Hu, S.-f. Fung \emph{et~al.}, ``Clustering social audiences in business information networks,'' \emph{Pattern Recognition}, vol. 100, p. 107126, 2020.

\bibitem{zhou2015cross}
X.~Zhou, X.~Liang, H.~Zhang, and Y.~Ma, ``Cross-platform identification of anonymous identical users in multiple social media networks,'' \emph{IEEE transactions on knowledge and data engineering}, vol.~28, no.~2, pp. 411--424, 2015.

\end{thebibliography}

%


\section{Biography Section}

\vspace{-30pt}
\begin{IEEEbiography}[{\includegraphics[width=1in,height=1.25in,clip,keepaspectratio]{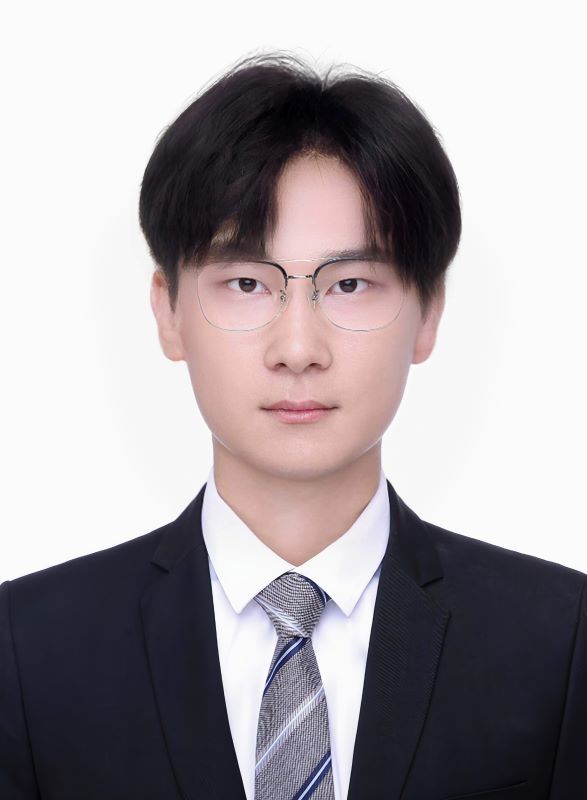}}]{Runzhen Xue}
received his B.E. degree from Shandong University, Qingdao, China in 2021. He is currently an M.S. candidate at Institute of Computing Technology, Chinese Academy of Sciences, Beijing, China. His current research interests include graph-based hardware accelerator and high-throughput computer architecture.
\end{IEEEbiography}

\vspace{-30pt}
\begin{IEEEbiography}[{\includegraphics[width=1in,height=1.25in,clip,keepaspectratio]{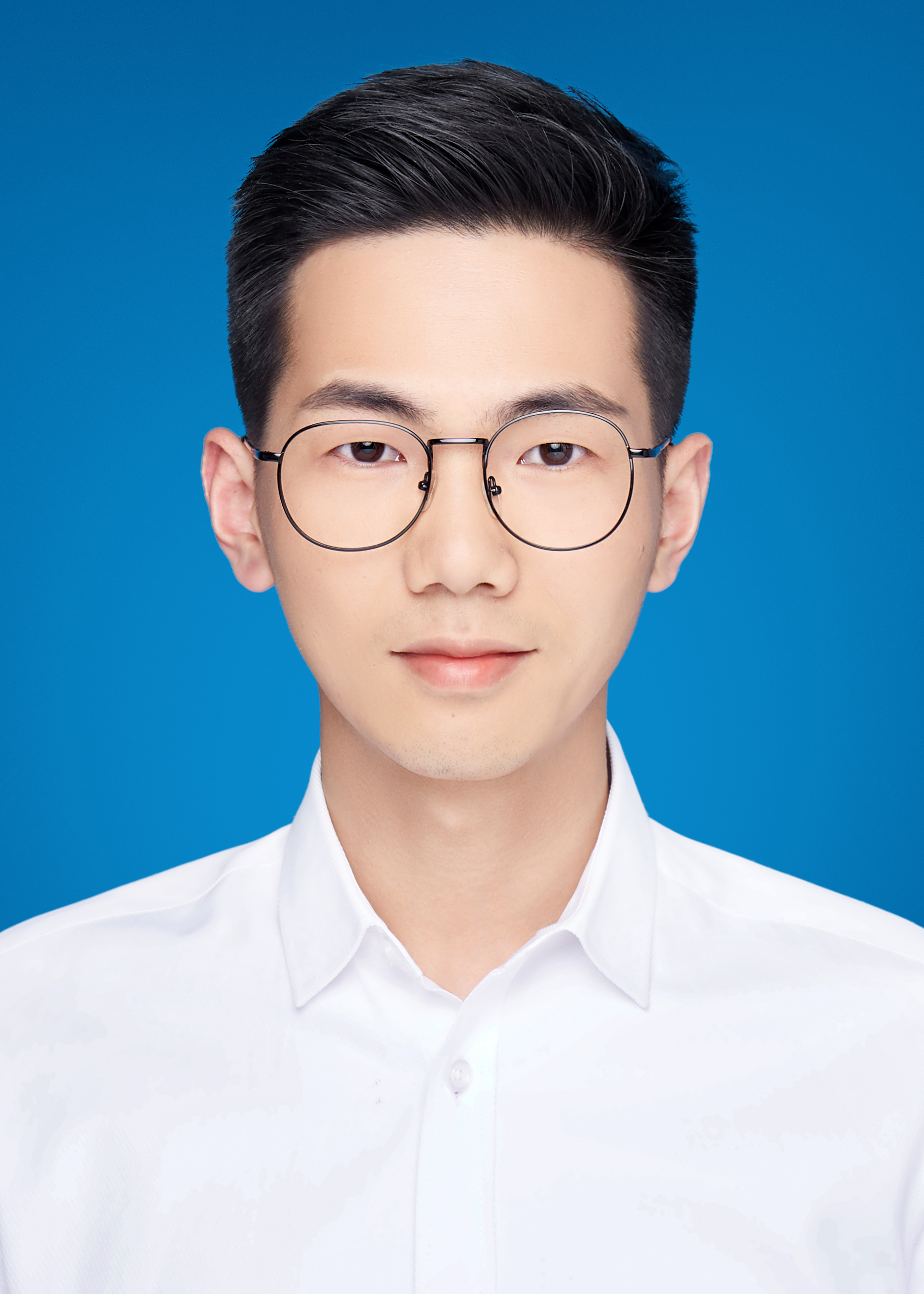}}]{Dengke Han}
He is currently an M.S. candidate at Institute of Computing Technology, Chinese Academy of Sciences, Beijing, China. His current research interests include graph-based hardware accelerator and high-throughput computer architecture.
\end{IEEEbiography}

\vspace{-30pt}
\begin{IEEEbiography}[{\includegraphics[width=1in, height=1.25in, clip, keepaspectratio]{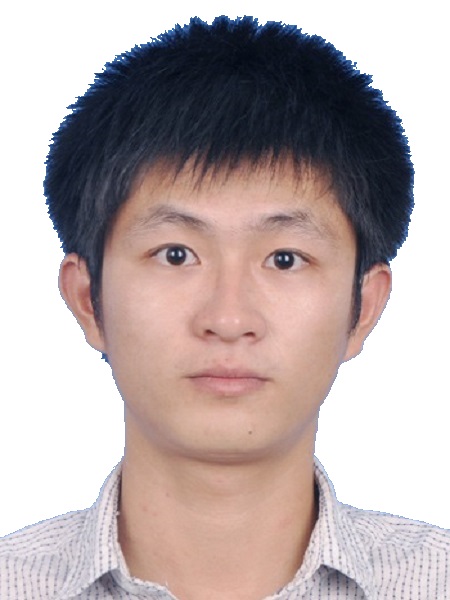}}] {Mingyu Yan} 
received his Ph.D. degree from University of Chinese Academy of Sciences, Beijing, China in 2020. He is currently an associate professor in Institute of Computing Technology, Chinese Academy of Sciences, Beijing, China. His current research interests include graph processing algorithm, graph-based hardware accelerator, and high-throughput computer architecture.
\end{IEEEbiography}

\vspace{-30pt}
\begin{IEEEbiography}[{\includegraphics[width=1in,height=1.25in,clip,keepaspectratio]{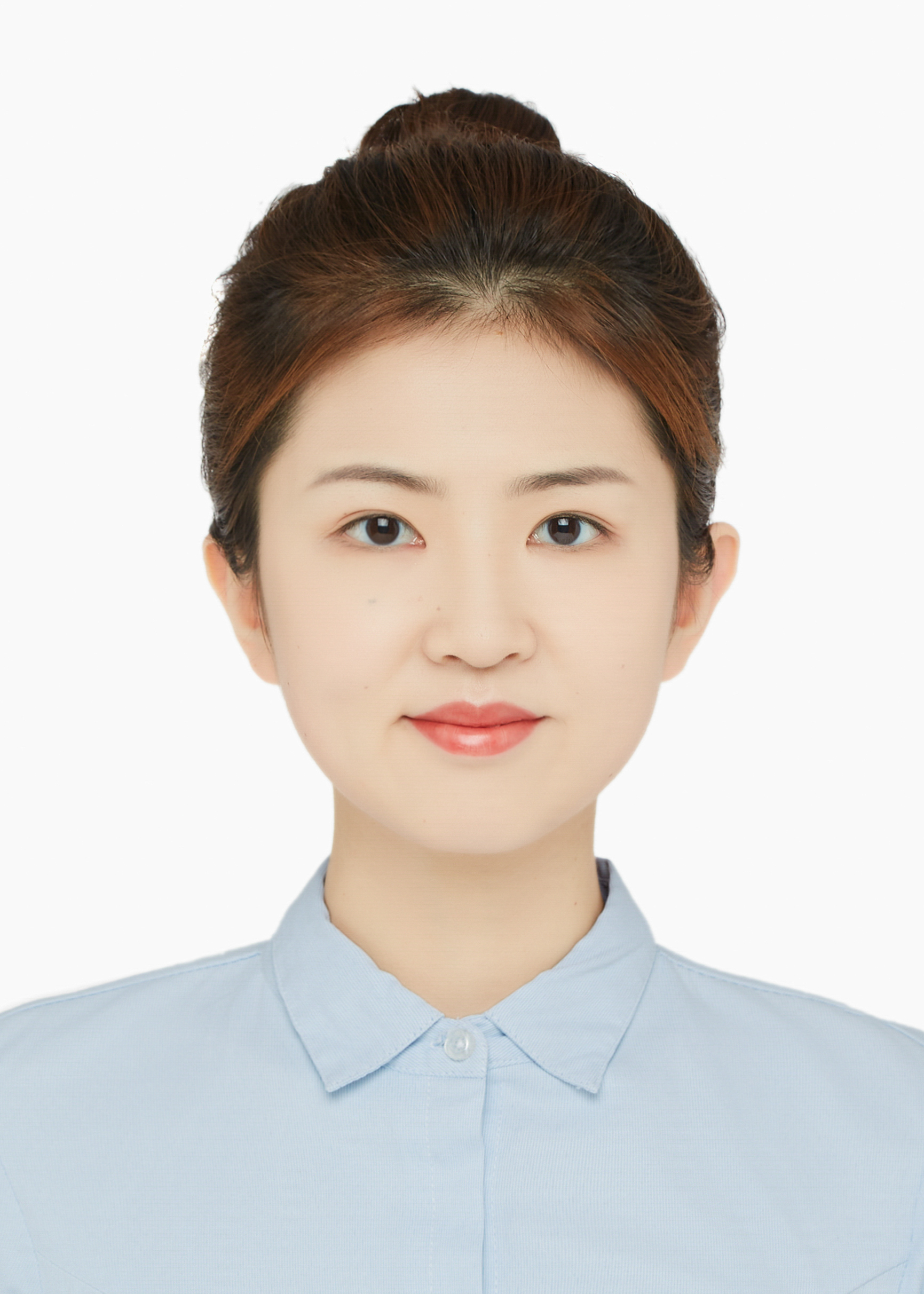}}]{Mo Zou}
received her Ph.D. degree from University of Chinese Academy of Sciences, Beijing, China in 2023. She is currently a postdoctoral researcher at Institute of Computing Technology, Chinese Academy of Sciences, Beijing, China. Her current research interests include computer architecture and memory system, especially on domain-specific hardware optimization.
\end{IEEEbiography}

\vspace{-30pt}
\begin{IEEEbiography}[{\includegraphics[width=1in,height=1.25in,clip,keepaspectratio]{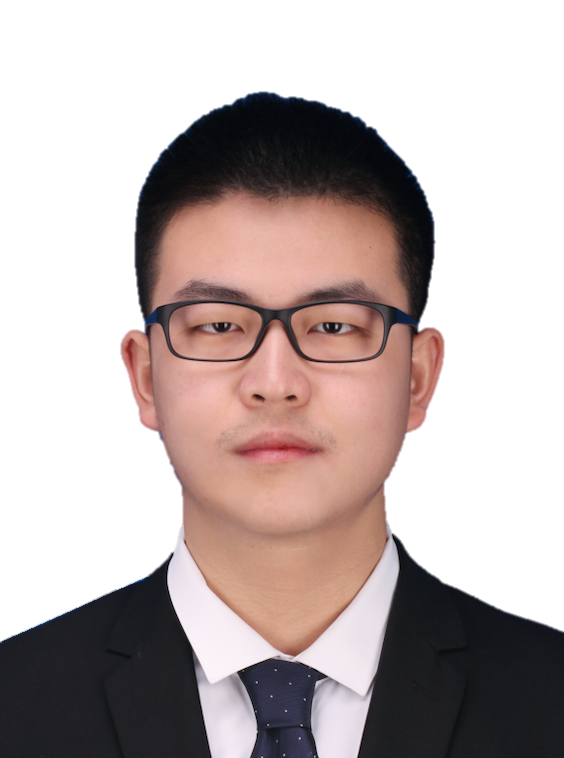}}]{Xiaocheng Yang}
received his M.S. degree from Tsinghua University, Beijing, China in 2020. He is currently a engineer at Institute of Computing Technology, Chinese Academy of Sciences, Beijing, China. His current research interests include machine learning algorithm and graph representation learning.
\end{IEEEbiography}

\vspace{-30pt}
\begin{IEEEbiography}[{\includegraphics[width=1in,height=1.25in,clip,keepaspectratio]{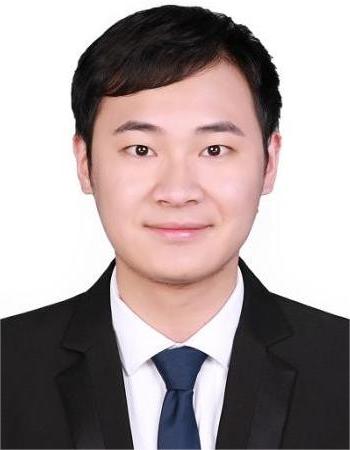}}]{Duo Wang}
received his B.S. degree from Southeast University, Nanjing, China in 2018. He is currently a Ph.D. candidate at Institute of Computing Technology, Chinese Academy of Sciences, Beijing, China. His current research interests include processor design space exploration, high-performance computer architecture and software simulation.
\end{IEEEbiography}

\vspace{-30pt}
\begin{IEEEbiography}[{\includegraphics[width=1in, height=1.25in, clip, keepaspectratio]{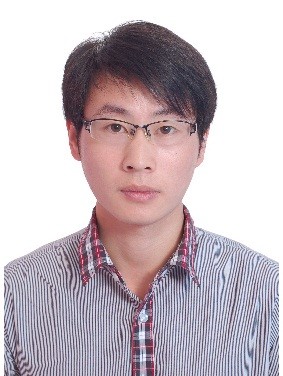}}] {Wenming Li} 
received the Ph.D. degree in computer architecture from Institute of Computing Technology, Chinese Academy of Sciences, Beijing, in 2016. He is currently an associate professor in Institute of Computing Technology, Chinese Academy of Sciences, Beijing. His main research interests include high-throughput processor architecture, dataflow architecture and software simulation. 
\end{IEEEbiography}

\vspace{-30pt}
\begin{IEEEbiography}[{\includegraphics[width=1in, height=1.25in, clip, keepaspectratio]{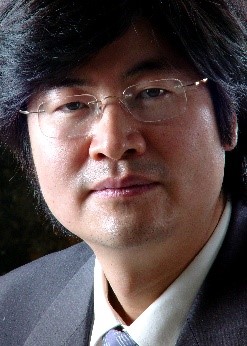}}] {Zhimin Tang} 
received his Ph.D. degree in computer architecture from Institute of Computing Technology, Chinese Academy of Sciences, Beijing, in 1990. He is currently a professor and Ph.D. supervisor in Institute of Computing Technology, Chinese Academy of Sciences, Beijing. His main research interests include high-throughput computer architecture, high-performance computer architecture and multi \& many-core processor design. 
\end{IEEEbiography}

\vspace{-35pt}
\begin{IEEEbiography}[{\includegraphics[width=1in, height=1.25in, clip, keepaspectratio]{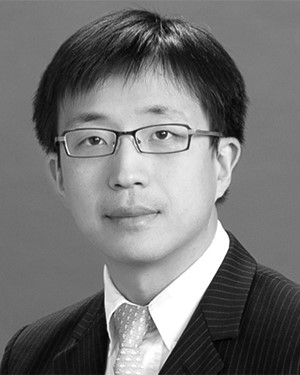}}] {John Kim}
received the BS and MEng degrees in electrical engineering from Cornell University, New York, in 1997 and 1998, respectively and the PhD degree in electrical engineering from Stanford University, in 2008. He is currently a Professor in the School of Electrical Engineering at KAIST.
His research interests include multicore architecture, interconnection networks, and datacenter architecture. 
\end{IEEEbiography}

\vspace{-35pt}
\begin{IEEEbiography}[{\includegraphics[width=1in, height=1.25in, clip, keepaspectratio]{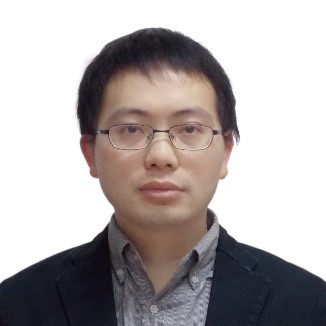}}] {Xiaochun Ye} 
received his Ph.D. degree in computer architecture from Institute of Computing Technology, Chinese Academy of Sciences, Beijing, in 2010. He is currently a professor and Ph.D. supervisor in Institute of Computing Technology, Chinese Academy of Sciences, Beijing. His main research interests include high-performance computer architecture and software simulation.
\end{IEEEbiography}

\vspace{-40pt}
\begin{IEEEbiography}[{\includegraphics[width=1in, height=1.25in, clip, keepaspectratio]{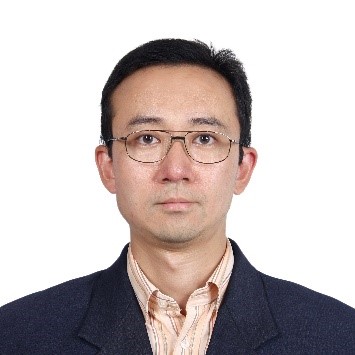}}] {Dongrui Fan} 
received his Ph.D. degree in computer architecture from Institute of Computing Technology, Chinese Academy of Sciences, Beijing, in 2005. He is currently a professor and Ph.D. supervisor in Institute of Computing Technology, Chinese Academy of Sciences, Beijing. His main research interests include high-throughput computer architecture and high-performance computer architecture. 
\end{IEEEbiography}

\vfill









\end{document}